%% file: access.tex
\documentclass{ieeeaccess}
\usepackage{cite}
\usepackage{amsmath,amssymb,amsfonts}
\usepackage{algorithmicx}
\usepackage{graphicx}
\usepackage{textcomp}
\def\BibTeX{{\rm B\kern-.05em{\sc i\kern-.025em b}\kern-.08em
    T\kern-.1667em\lower.7ex\hbox{E}\kern-.125emX}}

\usepackage{xcolor} 
\usepackage[labelfont=bf]{caption} 
\usepackage{textcomp} 
\usepackage{enumitem} 
\usepackage{lipsum} 
\usepackage{dblfloatfix} 

\usepackage{algorithm}
\usepackage{algpseudocode}

\algnewcommand\algorithmicinput{\textbf{Input:}}
\algnewcommand\Input{\item[\algorithmicinput]}

\algnewcommand\algorithmicconst{\textbf{Constraints:}}
\algnewcommand\Const{\item[\algorithmicconst]}

\algnewcommand\algorithmicoutput{\textbf{Output:}}
\algnewcommand\Output{\item[\algorithmicoutput]}

\algnewcommand{\algorithmicgoto}{\textbf{go to}}%
\algnewcommand{\Goto}[1]{\algorithmicgoto~\ref{#1}}%

\algrenewcommand\algorithmicindent{0.5em}

\usepackage[hidelinks, bookmarks=true]{hyperref}

\newlist{SubItemList}{enumerate}{1}
\setlist[SubItemList]{label={(\alph*)}}

\usepackage{array}
\newcolumntype{L}[1]{>{\raggedright\let\newline\\\arraybackslash\hspace{0pt}}m{#1}}
\newcolumntype{C}[1]{>{\centering\let\newline\\\arraybackslash\hspace{0pt}}m{#1}}
\newcolumntype{R}[1]{>{\raggedleft\let\newline\\\arraybackslash\hspace{0pt}}m{#1}}
\usepackage{makecell}
\usepackage{diagbox}

\let\OldItem\item
\newcommand{\SubItemStart}[1]{%
    \let\item\SubItemEnd
    \begin{SubItemList}[resume]%
        \OldItem #1%
}
\newcommand{\SubItemMiddle}[1]{%
    \OldItem #1%
}
\newcommand{\SubItemEnd}[1]{%
    \end{SubItemList}%
    \let\item\OldItem
    \item #1%
}
\newcommand*{\SubItem}[1]{%
    \let\SubItem\SubItemMiddle%
    \SubItemStart{#1}%
}%

\newcommand{\changed}[1]{\textcolor{black}{#1}}

\newlist{inlinelist}{enumerate*}{1}
\setlist[inlinelist]{label=(\arabic*)}

\begin{document}

\history{Date of publication xxxx 00, 0000, date of current version xxxx 00, 0000.}
\doi{10.1109/ACCESS.2017.DOI}

\title{Architectural-Space Exploration of Heterogeneous Reliability and Checkpointing Modes for Out-of-Order Superscalar Processors}
\author{\uppercase{Bharath Srinivas Prabakaran}\authorrefmark{1},
\IEEEmembership{Student Member, IEEE}, 
\uppercase{Mihika Dave}\authorrefmark{2}, 
\uppercase{Florian Kriebel\authorrefmark{1}},
\uppercase{Semeen Rehman\authorrefmark{1}, and} 
\uppercase{Muhammad Shafique\authorrefmark{1}},
\IEEEmembership{Senior Member, IEEE}}
\address[1]{Technische Universit{\"a}t Wien (TU Wien), Vienna, Austria}
\address[2]{University of Illinois Urbana-Champaign, Champaign, Illinois, USA}
\tfootnote{This work is supported in parts by the German Research Foundation (DFG) as part of the priority program "Dependable Embedded Systems" (\href{http://spp1500.itec.kit.edu}{SPP 1500 - http://spp1500.itec.kit.edu})}

\markboth
{Prabakaran \headeretal: Heterogeneous Reliability and Checkpointing Modes for Out-of-Order Superscalar Processors}
{Prabakaran \headeretal: Heterogeneous Reliability and Checkpointing Modes for Out-of-Order Superscalar Processors}

\corresp{Corresponding author: Bharath Srinivas Prabakaran (\href{mailto:bharath.prabakaran@tuwien.ac.at}{bharath.prabakaran@tuwien.ac.at}).}

\begin{abstract}
Reliability for multi-core processors has emerged as an important design constraint.
A key research challenge is to detect and/or mitigate \textit{transient faults}, such as soft errors, that can abruptly terminate an executing application or generate incorrect output, both leading to undesirable effects that can potentially be catastrophic in safety-critical systems.
State-of-the-art reliability techniques and mechanisms deploy full-scale redundancy, like double or triple modular redundancy (DMR, TMR), on different layers of the computing stack to detect and/or correct such transient faults.
However, the techniques relying on full-scale redundancy incur significant area, performance, and/or power overheads, which might not always be feasible/practical due to system constraints such as deadlines and available power budget for the full chip (or a processor core).
Moreover, depending on the inherent resilience of an application, not every application requires full-scale redundancy, that would, otherwise, result in resource/energy wastage.
Hence, techniques relying on selective redundancy have recently been investigated by researchers.

\quad In this work, we propose a novel design methodology to generate and explore the architectural-space of heterogeneous reliability modes for out-of-order superscalar multi-core processors.
These heterogeneous modes are iso-ISA (i.e., implement the same Instruction Set Architecture), but differ in terms of the micro-architectural implementation, i.e., different components are hardened with different reliability techniques.
Hence, these heterogeneous modes enable varying reliability and power/area trade-offs, from which an optimal configuration can be chosen at run time to meet the reliability requirements of a given system, while reducing the corresponding power overheads (or alternatively solving the inverse problem, i.e., maximizing the reliability under a given power constraint).
We implemented different reliability modes for the ALPHA 21264 out-of-order superscalar microprocessor, and integrated different cores with heterogeneous reliability modes in a multi-core configuration.
Our experimental results show that a pareto-optimal heterogeneous reliability mode reduces the core vulnerability by 87\%, on average, across multiple application workloads, with area and power overheads of 10\% and 43\%, respectively.

\quad To further enhance the design space of heterogeneous reliability modes, we investigate the effectiveness of combining different processor state compression techniques like Distributed Multi-threaded Checkpointing (DMTCP), Hash-based Incremental Checkpointing (HBICT) and GNU zip, such that the correct processor state can be recovered once a fault is detected.
These state compression techniques aim at reducing the storage requirements of the processors' correct state, which is backed-up at an application checkpoint during its execution to ensure successful recovery.
We reduced the checkpoint sizes by a factor of \textasciitilde6$\times$ using a unique combination of different state compression techniques. 
To validate our concepts, we significantly enhanced the open-source cycle-accurate simulator \textit{gem5} (which is widely adopted by the relevant research communities) with the state compression techniques and the heterogeneous reliability modes.
\end{abstract}


\titlepgskip=-20pt

\maketitle

\input{sections/section1.tex}

\input{sections/section2.tex}
\input{sections/section3.tex}
\input{sections/section4.tex}
\input{sections/section5.tex}
\input{sections/section6.tex}

\bibliographystyle{IEEEtran}
\bibliography{References}

\input{sections/z_biography.tex}

\EOD

\end{document}

%% file: sections/section1.tex
\section{Introduction}
\label{sec:Intro}

\IEEEPARstart{A}{ggressive} transistor scaling has led to an increased susceptibility towards several reliability problems, such as soft errors, at the hardware layer~\cite{henkel2013reliable}. 
Soft errors are \textit{transient faults} in the hardware that cause bit-flips in the micro-architecture, which may propagate to the application output and corrupt its state, or may terminate the application's execution~\cite{baumann2005radiation}~\cite{SoftErrorStudy}. 
The rate of occurrences of these soft errors is expected to increase with each new generation of microprocessors being released, due to aggressive shrinking of transistors' feature sizes and imperfection in the fabrication process~\cite{feng2010shoestring}~\cite{BorkarMicro} (see Section~\ref{sec:Preliminaries}).

Plenty of research works focusing on techniques like full-scale redundancy and checkpointing have been proposed towards prevention, detection, and/or mitigation of soft errors across the computing stack, i.e., the hardware and software layers~\cite{li2012reli}~\cite{li1990catch}. 
Reliability at the hardware layer is ensured through redundancy of execution paths and/or hardening of pipeline components, i.e., full-scale Double or Triple Modular Redundancy (DMR, TMR). 
Software-layer techniques realize full-scale spatial/temporal redundancy by executing multiple redundant instructions or threads of an application, thereby ensuring a reliable output~\cite{mukherjee2002detailed}\cite{OhTREDDI}\cite{ReisSWIFT}. 
However, these full-scale redundancy techniques incur significant performance and energy overheads (e.g., in case of temporal redundancy), and area/power/energy overhead (e.g., in case of spatial redundancy). 

Therefore, we propose to investigate the individual properties and requirements of an application workload to determine the component-level vulnerabilities of an out-of-order superscalar processor at design-time, i.e., enabling reliability provisions at a much finer granularity. 
Based on this analysis, we develop a \textit{wide range of heterogeneous reliability and checkpointing modes} that enable efficient control over the achieved reliability and the incurred overhead, especially when considering the diverse resilience properties of different executing applications at different run time instances.
Our previous work~\cite{rehman2018hardware} provides an initial proof-of-concept of this work and preliminary results for the feasibility of reliability-heterogeneous cores.
In this work, we significantly extend this concept, and provide a systematic methodology to integrate such reliability heterogeneous modes on a chip along with other different types of reliability mechanisms like checkpointing and state compression to expand the space of design trade-offs for reliability vs. overhead.

\begin{figure}[h]
    \centering
    \captionsetup{justification=raggedright,singlelinecheck=false}
    \includegraphics[width = \linewidth]{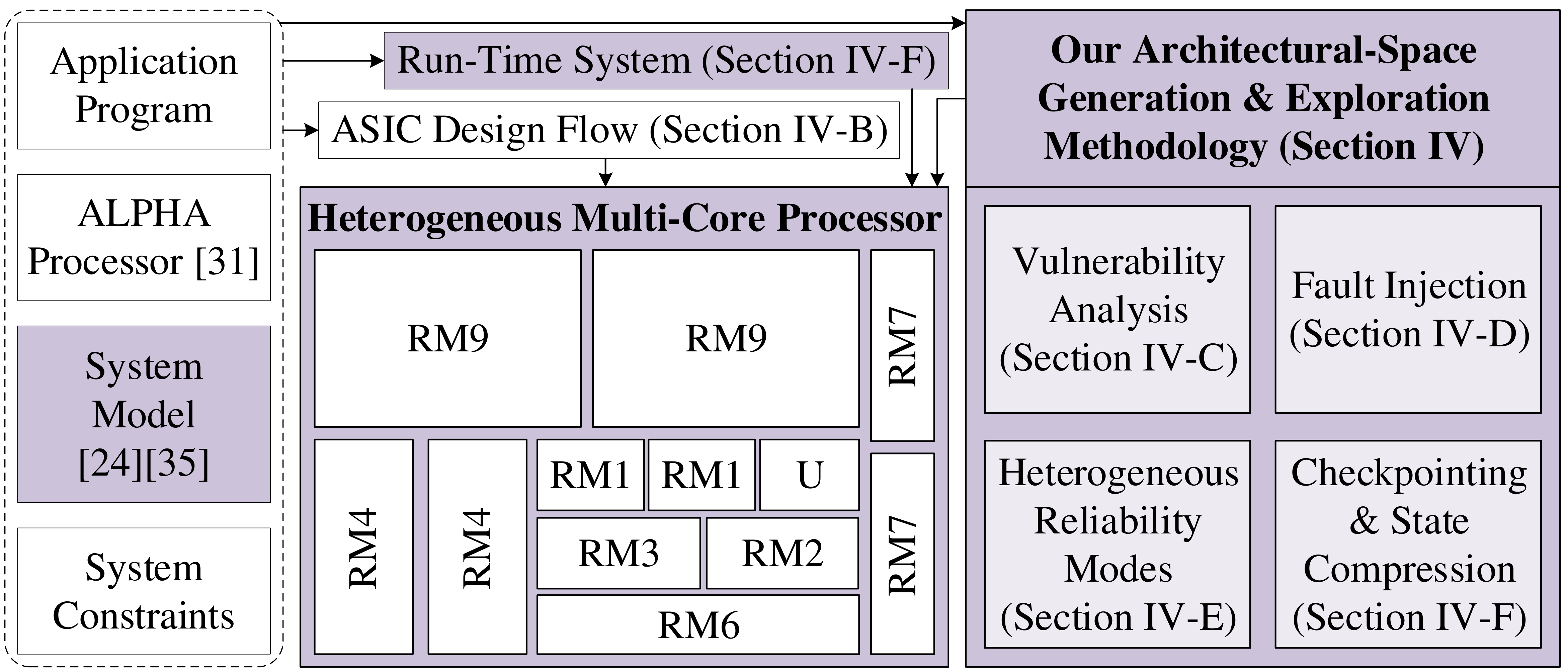}
    \caption{\textbf{An Overview of our Contributions (Highlighted Boxes) in the Processor Design Flow.}}
    \label{fig:DF}
\end{figure}

In a nutshell, \textbf{we make the following novel contributions:}
\begin{enumerate}[leftmargin=*,label={(\arabic*)}]
    \item \textit{Vulnerability Analysis:} A comprehensive vulnerability analysis of an out-of-order superscalar core to compute the Architectural Vulnerability Factor (AVF) of different pipeline components.
    \item \textit{Methodology for Architectural-Space Generation and Exploration:} A novel architectural-space generation and exploration methodology that:
    \begin{enumerate}[label={(\alph*)}]
        \item analyzes the processor-level vulnerabilities of out-of-order superscalar cores to develop a wide range of heterogeneous reliability modes at design-time;
        \item provides a multi-core processor with multiple heterogeneous reliability modes, such that each core deploys distinct reliability measures in different components;
        \item enables the reliability-power trade-offs of the proposed heterogeneous reliability modes under diverse application workloads, which can be leveraged at run-time to either \textit{optimize reliability under the given power constraint}, or \textit{decrease power consumption under the reliability requirement of the system};
    \end{enumerate}
    \item \changed{\textit{Run-Time System:} To evaluate the run-time benefits of reliability-heterogeneity, we execute multiple different application workload mixes on our heterogeneous 10-core processor with all the designed heterogeneous reliability modes.
    We propose and evaluate two different mapping policies, namely, Vulnerability-Constrained Power Minimization and Power-Constrained Vulnerability Minimization.
    These two policies differ in terms of the constraints imposed and their minimization objectives, as provided by the system designer.
    We illustrate the decrease in power overhead and the full-processor vulnerability factor using these two task-mapping policies.
    }
    \item \textit{State Compression:} To further enhance the processor reliability and to increase the design space, we analyze and investigate efficient state compression techniques. 
    These strategies can be used to decrease the storage size of the checkpointing data, for example, in case of \textit{transient faults}, and to enable an efficient rollback of the processor to its last known safe-state. 
    \item \textit{Evaluation and Discussion:} We evaluate the effectiveness of the heterogeneous reliability modes designed using our proposed methodology under diverse application workloads by enhancing the widely adopted cycle-accurate simulator \textit{gem5} to offer the required functionality of heterogeneous-reliability and state compression. 
\end{enumerate}
Fig.~\ref{fig:DF} illustrates an overview of our contributions in a design-flow for developing heterogeneous multi-core processors.

\textbf{Paper Organization:}
Section~\ref{sec:Preliminaries} presents the preliminaries and background information required to understand our proposed contributions. We discuss the system models in Section~\ref{sec:SystemModel}.
Section~\ref{sec:HHM} presents our methodology for generation and exploration of the architectural-space of heterogeneous reliability modes, including results that illustrate the benefits of the proposed approaches.
Section~\ref{sec:RW} presents the related work on state-of-the-art reliability techniques and heterogeneous reliability approaches, followed by the conclusion presented in Section~\ref{sec:conclusion}.

%% file: sections/section2.tex
\section{Preliminaries and Background}
\label{sec:Preliminaries}
\textbf{Soft Errors:} In the era of nanometer technology nodes, reliability threats like manufacturing-induced \textit{process variation}, \textit{device aging}, and \textit{transient faults} are increasingly challenging the functional correctness and safety-critical aspects of the systems where these electronic devices are deployed~\cite{henkel2013reliable}.
An example of \textit{transient faults} is \textbf{soft errors}, which have emerged as a serious threat to the reliability of a digital system. These soft errors are generated at the hardware layer, due to four key factors, namely,
\begin{enumerate}[leftmargin=*,label=(\arabic*)]
\item \textit{Alpha Particles}, which are positively charged composite particles emitted during radioactive decay. These particles travel through the semiconductor device thereby disturbing the electron distribution of the transistor~\cite{MayAlphaParticles}.
\item \textit{Cosmic Rays}, which are a flux of energetic neutrons that are constantly emitted by the solar system~\cite{SrinivasaIRPS}\cite{GormanTED}.
\item \textit{Thermal Neutrons}, which are neutrons that have attained thermal equilibrium after dissipating all kinetic energy~\cite{HazuchaTNS}.
\item \textit{Internal Factors} such as random noise, signal integrity issues, cross-talks, and electromagnetic interference~\cite{SoftErrorStudy}.
\end{enumerate}

\begin{figure}[t]
    \centering
    \captionsetup{justification=raggedright,singlelinecheck=false}
    \includegraphics[width = 0.9\linewidth]{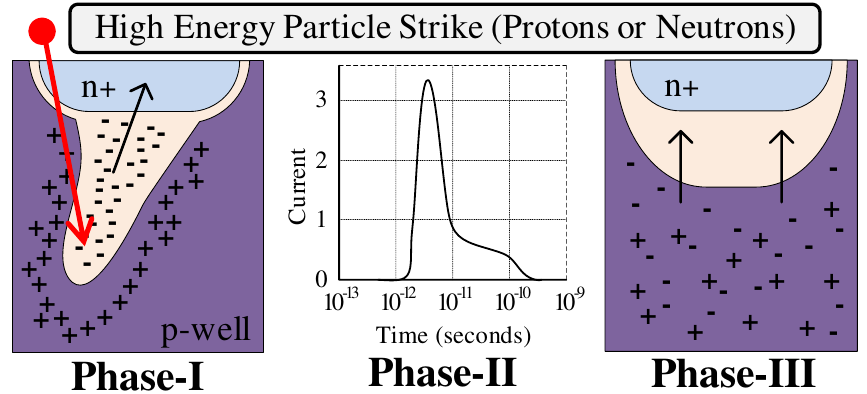}
    \caption{\textbf{The Three Phases of the Soft Error\hspace{\textwidth} Phenomenon~(adapted from~\cite{baumann2005radiation}).}}
    \label{fig:SE}
\end{figure}

Soft errors cause temporary bit-flips either in the control or data path of a micro-architecture, or in the on-chip memory cells, which may propagate to the application output or may crash, hang, or terminate the application execution~\cite{baumann2005radiation}.
Fig.~\ref{fig:SE} illustrates the soft-error phenomenon, which can be broken into three phases. 
\begin{enumerate} [leftmargin=*,label=(\arabic*)]
    \item First, in the \textit{ion-track formation} phase (phase-I), a high energy particle (such as the cosmic rays discussed earlier) strikes the transistor to generate multiple electron-hole pairs, which in turn increase the concentration of carriers along the ion's path.
    \item In phase-II (\textit{current pulse generation}), the ions collected at the depletion region form a ``temporary'' channel that funnels the current from source to drain, which could toggle the transistor state for tens of picoseconds.
    This can result in a bit flip in
    \begin{inlinelist}[leftmargin=*,label=(\roman*)]
    \item the memory cell, which can be latched to the incorrect value until and unless its is over-written by another value; or
    \item the logic gate that can potentially propagate to the final output of the circuit, thereby corrupting the output of the circuit.
    \end{inlinelist}
    \item In the \textit{ion diffusion} phase (phase-III), over a period of tens or hundreds of picoseconds, the charges diffuse into the depletion layer, thereby disintegrating the temporary channel.
\end{enumerate}

\begin{figure}[h]
    \centering
    \captionsetup{justification=raggedright,singlelinecheck=false}
    \includegraphics[width = \linewidth]{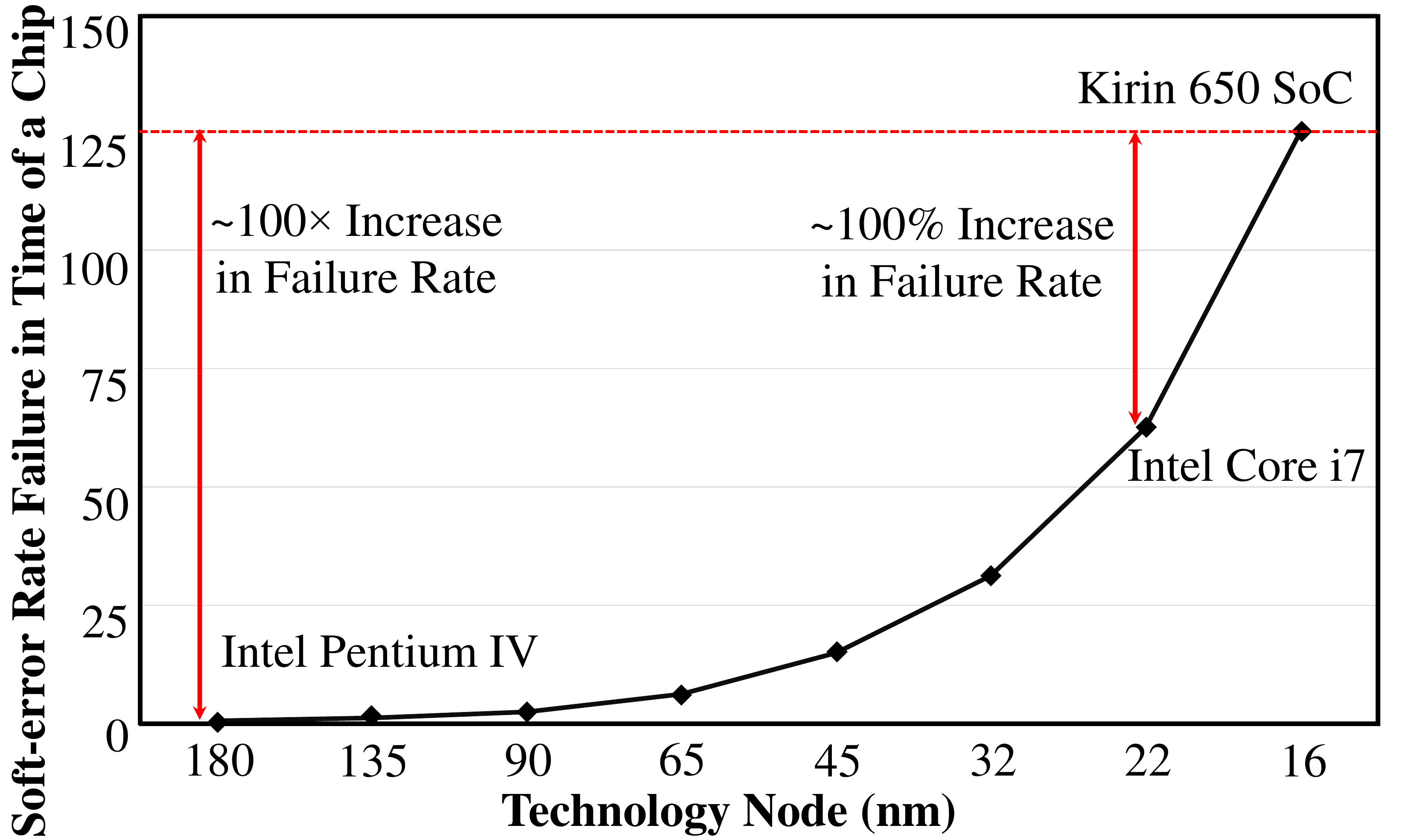}
    \caption{\textbf{Increase in Soft-error Rate of a Chip for Multiple Technology Nodes (adapted from~\cite{BorkarMicro}).}}
    \label{fig:SER}
\end{figure}

\textbf{Increasing Soft Error Rates:} In the earlier generation technology nodes, the transistor dimensions were large enough that a temporary channel could not funnel the current from source to drain.
Furthermore, due to reducing transistor dimensions, the rate of soft error occurrences is increasing with each new generation of processors being released into the market, due to their fabrication using continuously smaller technology nodes ~\cite{feng2010shoestring}\cite{BorkarMicro} (see Fig.~\ref{fig:SER}).
\changed{This is a major threat for the current world infrastructure, which heavily relies on electronics for all activities, such as work, communication, transportation, socializing, internet, etc.
Even the day-to-day devices and services that people use, e.g., wearable devices such as smart-watches and fitness trackers, mobile computing platforms such as mobile phones and laptops, and on-demand cloud services offered by large-scale data centers, heavily rely on the reliability of electronic devices.
This becomes even more crucial for safety-critical application domains like aerospace, automotive, healthcare, industry 4.0, smart grids, smart homes, etc.}

\begin{figure}[h]
    \centering
    \captionsetup{justification=raggedright,singlelinecheck=false}
    \includegraphics[width = \linewidth]{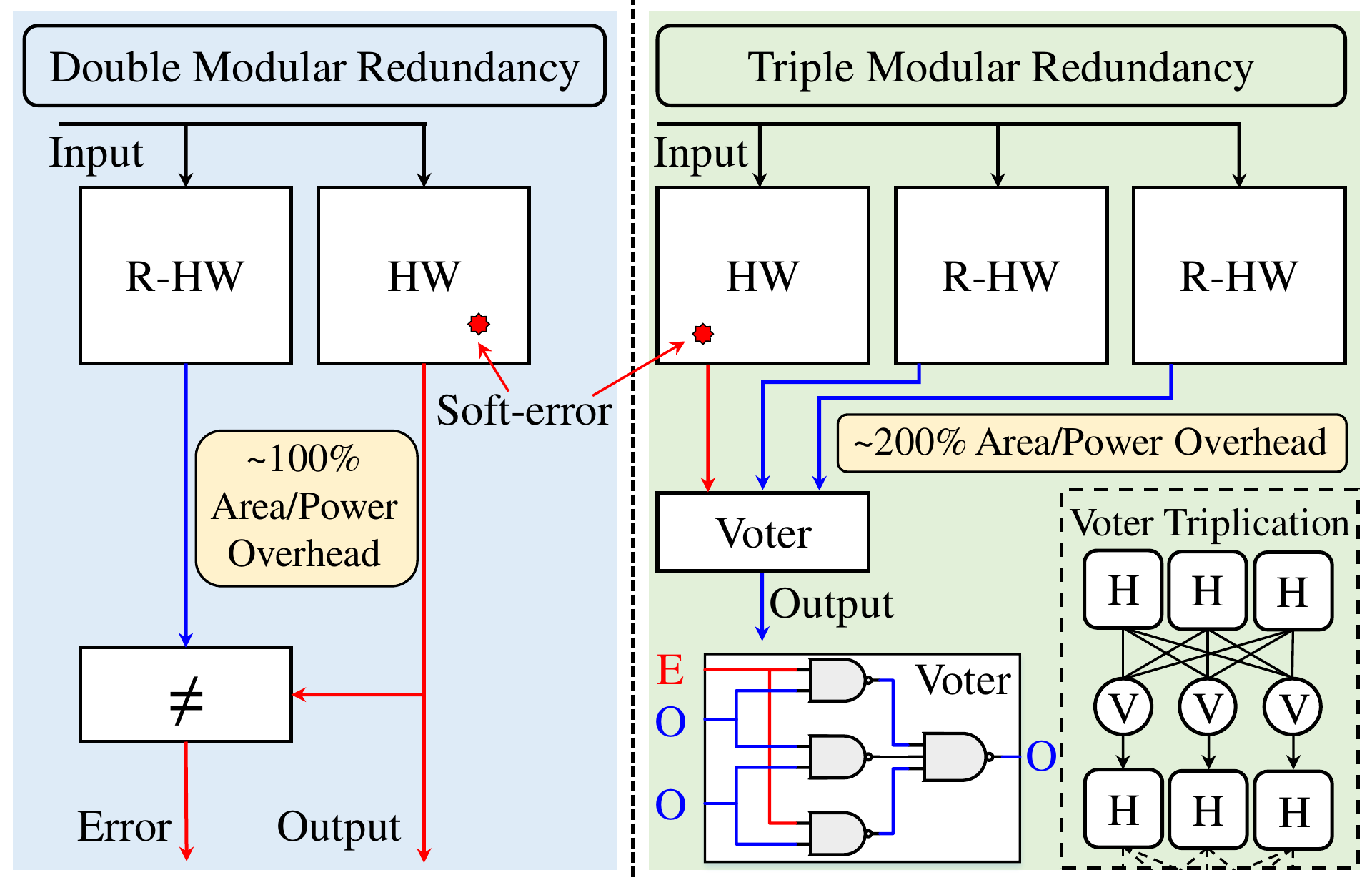}
    \caption{\textbf{An Overview of the Redundancy Techniques at the Hardware Layer.}}
    \label{fig:Hardening}
\end{figure}

\textbf{Processor Hardening:} Reliability at the hardware layer is \changed{typically} ensured by the use of \changed{full-scale} redundancy, which involves realizing multiple instantiations of the hardware unit with the same set of inputs, to generate outputs that can be compared with each other to detect errors (in DMR) or correct errors using a voter circuit (in case of TMR), which we refer to as \textit{hardware hardening}. 
An overview of these hardware-level redundancy techniques is presented in Fig.~\ref{fig:Hardening}.
DMR and TMR incur significant area and power overheads caused by the redundant hardware units and the additional circuitry used to detect or correct errors.
Furthermore, since the additional hardware components execute in parallel, the throughput of the system is not affected, with a minimal gate-level increase in delay caused by the voter circuit.
Typically, to ensure very high reliability, the entire processor pipeline (full-scale) is hardened, i.e., all the pipeline components are instantiated thrice with the same set of inputs and a voter circuit to elect the majority output\changed{, as illustrated by Gaisler's completely hardened LEON3-FT micro-processor that deploys redundancy in the register file and cache memory~\cite{LEON3FT}.}
Fig.~\ref{fig:Hardening} also illustrates the gate-level implementation of the voter circuit, and how, in the case of soft errors, the majority output is elected and generated as the final output.
Note, this leads to the possibility of the voter circuit becoming a single point of failure, which is mitigated by triplicating the voter circuit as well, and has been deployed, for example, in the Saturn Launch Vehicle Digital Computer~\cite{IBM:1962}\cite{SaturnV}.
In this work, \changed{without the loss of generality, we advocate the enabling of fine-grained reliability at different component level that can facilitate realization of different hardening modes for different processor cores, thereby providing a wide range of reliability-power trade-offs. 
As a proof of concept, we will showcase an example of using component-level TMR with a single majority voter circuit.
However, any other reliability mechanism can be deployed as a knob at the component level.}

\begin{figure}[t]
    \centering
    \captionsetup{justification=raggedright,singlelinecheck=false}
    \includegraphics[width = \linewidth]{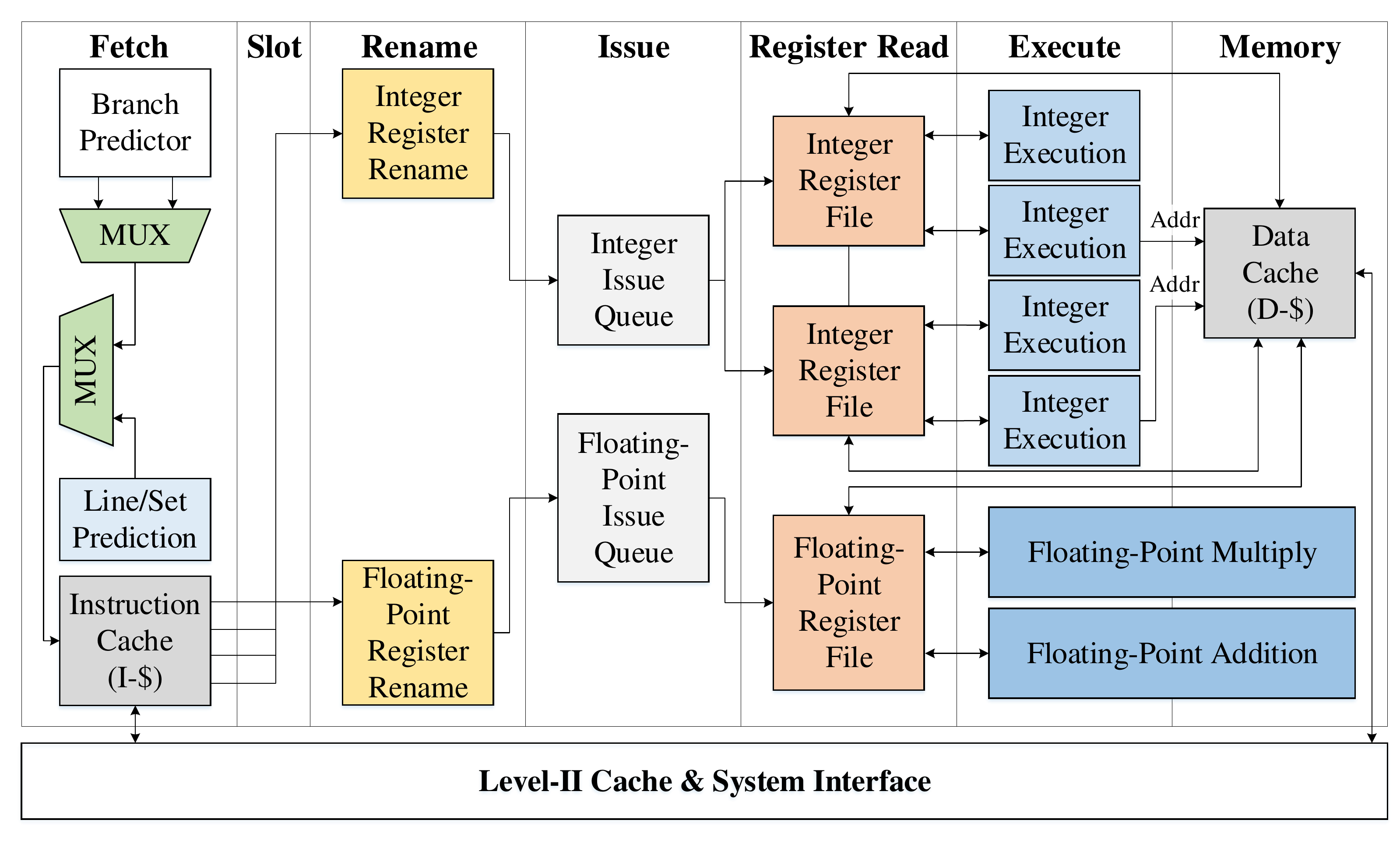}
    \caption{\textbf{ALPHA 21264 Out-of-Order Superscalar Processor Architecture (adapted from~\cite{IEEE:ALPHA}).}}
    \label{fig:Alpha}
\end{figure}

\textbf{Out-of-Order Superscalar Processors:}
Besides transistor scaling, architectural innovations such as deep pipelining, instruction-level parallelism, out-of-order execution, speculative execution, branch prediction, etc. have tremendously increased the computing capabilities of microprocessors. 
Almost all current generation microprocessors are realized with such functionalities to ensure high system performance.
For example, \textit{superscalar processors} exploit an application's instruction-level parallelism to execute multiple instructions in parallel during the same clock-cycle on multiple different execution units~\cite{Johnson:Superscalar}.
\textit{Out-of-Order processors} execute instructions out-of-order, as opposed to the typical sequential execution, by exploiting the interdependency, or the lack thereof, of program instructions and the data processed by them~\cite{Patt:OOO}.
This allows for executing ``independent'' instructions in clock-cycles that would be otherwise lost in pipeline stalls caused by control- or data-flow dependencies.
Fig.~\ref{fig:Alpha} illustrates the control- and data-path of the ALPHA 21264 out-of-order superscalar microprocessor~\cite{IEEE:ALPHA}, \changed{which is widely used in the architecture research community.}

Alpha 21264, or Alpha 7, is a four-issue, seven pipeline stage superscalar processor architecture that is capable of executing up to six (four integer and two floating-point) instructions per cycle (IPCs) while sustaining four instructions simultaneously.
During a program's execution, the processor can accommodate up to 80 instructions in the pipeline, which is kept track of using the processor's re-order buffer (ROB).
The Alpha 7 processor also includes two cache levels, i.e., the primary and secondary caches. 
The processor uses a modified Harvard architecture that implements separate primary instruction (I-cache) and data caches (D-cache), typically of size 64KB each.
The D-cache is dual-ported in order to allow simultaneous read and write on both rising and falling edge of the clock.
This feature allows for reducing the area and power overheads associated with duplicating the cache, as in the Alpha 21164 microprocessor.
The secondary cache, or B-cache, is usually a direct-mapped cache that is located off-chip and shared by all processor cores.
Typically, L2-cache has a maximum capacity of 16MB and is constructed using synchronous static random access memory (SSRAM), which is accessed using a dedicated 128-bit high-bandwidth bus~\cite{CompaqALPHA}.
Branch prediction in this microprocessor is implemented using a hybrid two-level branch prediction algorithm called tournament prediction, with a minimum branch misprediction penalty of $7$ clock-cycles~\cite{CompaqALPHA}.
The processor was built using 15.2 million transistors, roughly $40\%$ of which was occupied by the core processing unit and the rest of which was consumed by the caches and branch history tables~\cite{GronowskiJSSC}.



%% file: sections/section3.tex
\section{System Model}
\label{sec:SystemModel}
\textbf{Architecture Model:}
\changed{
To cater for different application workloads with varying reliability requirements, we realize a reliability-heterogeneous multi-core processor ($HMC$):
\begin{center}
    $HMC=\{PC_1, PC_2, ..., PC_M\}$
\end{center}
where $PC_j$ denotes the $j^{th}$ processor core, such that, $j \in \{1, 2, ..., M\}$, with a total of $M$ processor cores in the $HMC$.
Each processor core has $L$ different architectural components, denoted as:
\begin{center}
    $PC_j=\{C_{(j,1)}, C_{(j,2)}, ..., C_{(j,L)}\}$
\end{center}
where $C_{(j,k)}$ denotes the $k^{th}$ component in the $j^{th}$ processor core. 
Each architectural component (like re-order buffer, register file, instruction queue, etc.) in each processor core ($C_{(j,k)}$) can be hardened by using mechanisms like TMR, DMR, Checkpointing and Rollback, Error Correcting Codes, or Razor latches. 
We denote the $i^{th}$ reliability technique of the component $C_{(j,k)}$ as:
\begin{center}
    $RT(C_{(j,k)}) = i$
\end{center}
Without loss of generality, in this work, we explore the applicability of TMR for designing the heterogeneous reliability modes.
This leads to $ i = \{0,1\}$, where $RT(C_{(j,k)}) = 0$ denotes the unprotected component without any type of hardening and $RT(C_{(j,k)}) = 1$ denotes a component that has been hardened by triple modular redundancy, thereby enabling heterogeneous hardening.
}


The area of each processor core is denoted as $A(PC_j)$, which is the summation of area of all the processor components, including the overhead of hardening certain components.
%
%
\changed{Note, only a selective subset of the different heterogeneous reliability modes can be activated at run-time due to the total power constraint of a system, while considering the application's reliability requirement.
An overview of the symbols used in this work and their denotations have been presented in Table~\ref{table5}.}

\begin{table}[h]
	\centering
	\caption{\protect\centering\textbf{Symbols and Denotations}}
	\begin{tabular}{|C{2cm}|L{4.6cm}|}
		\cline{1-2}
		\textbf{Symbol} & \textbf{Denotation} \\ \hline
		$HMC$ & Heterogeneous Multi-core Processor \\ \hline
		$M$ & Total Number of Cores in $HMC$ \\ \hline
		$PC_j$ & $j^{th}$ Processor Core \\ \hline
		$C_{(j,k)}$ & $k^{th}$ Architectural Component \newline in $j^{th}$ Processor Core \\ \hline
		$L$ & Total Number of Architectural Components in $PC_j$ \\ \hline
		$RT(C_{(j,k)})$ & The Reliability Technique used \newline to Harden Component $C_{(j,k)}$ \\ \hline
		$A(PC_j)$ & Area of the Processor Core $PC_j$ \\ \hline
		$T$ & Set of Tasks \\ \hline
		$Z$ & Total Number of Tasks \\ \hline
		$E$ & Set of Task Dependencies \\ \hline
		$P(T_q, PC_j)$ & Power Consumption for the $q^{th}$ Task when executing on $j^{th}$ Processor Core \\ \hline
		$L(T_q, PC_j)$ & Average Execution Time for the $q^{th}$ Task when executing on $j^{th}$ Processor Core \\ \hline
		$AVF_{C_{(j,k)}}$ & Architectural Vulnerability Factor\newline of Component $C_{(j,k)}$ \\ \hline
		$FPVF(T_q, PC_j)$ & Full-Processor Vulnerability Factor of Processor Core $PC_j$ for Task $T_q$ \\ \hline
		$N$ & Number of Clock Cycles \\ \hline
		$VulnerableBits$ & Total Number of Vulnerable Bits \\ \hline
		$VulnerableTime$ & Time Duration of Vulnerable Bits \\ \hline
		$TotalBits$ & Total Number of Output Bits \\ \hline
		$TotalTime$ & Total Duration of Application Execution \\ \hline
		$P_{error}$ & Error Rate of a Transient Fault \\ \hline
		$P_{flip}$ & Probability of High-Energy Particle Strike leading to a Soft Error \\ \hline
		$N_{error}$ & Number of Program Failures \\ \hline
		$N_{FI}$ & Spatial Vulnerability of Component \\ \hline
		$ORM$ & Optimal Reliability Modes \\ \hline
		\end{tabular}
	\label{table5}
\end{table}

\textbf{Application Model:}
The applications are modeled as a set of task graphs $\{T,E\}$ containing task and dependency information for all application workloads.
$T$ is denoted as $T=\{T_1, T_2, ..., T_Z\}$ for a set of $Z$ tasks. $E$ is defined as $E=\{E_{xy} \mid (T_x,T_y) \in T\}$ for the set of task dependencies.
For the given processor core ($PC_j$) each task $T_q$ has the following execution properties: 
\begin{itemize}[leftmargin=*]
    \item $P(T_q,PC_j)$, which denotes the peak power consumption,
    \item $L(T_q,PC_j)$, which denotes the average performance in terms of execution time, and
    \item $FPVF(T_q,PC_j)$, which denotes the full-processor vulnerability factor.
\end{itemize}

\textbf{Reliability Model:}
The Architectural Vulnerability Factor (AVF) of a hardware component is defined as the probability of a fault to propagate to the final output resulting in an execution error~\cite{1253181}. 
We compute the $AVF$ of a component $C_{(j,k)}$ as the fraction of bits vulnerable in each cycle (\textit{Vulnerable-Bits}) to the total number of output bits (\textit{TotalBits}) generated by component $C_{(j,k)}$ for a duration of $N$ cycles. 
AVF of a component $C_{(j,k)}$ is `0' if the component is hardened, or produces no architecturally incorrect bits~\cite{1253181}. 
Note, all bits of a branch predictor are always architecturally correct, therefore a branch predictor's AVF is always `0'. Similarly, all bits of the program counter (PC) are always vulnerable, therefore the AVF of a PC is always `100'~\cite{1253181}.
AVF is estimated using the following equation:

\begin{center}
$AVF_{C_{(j,k)}} = \frac{\sum_{n=0}^{N}VulnerableBits(C_{(j,k)})}{TotalBits \times N}\times 100$
\end{center}

To study the impact of component hardening on the full-processor, we extend the $AVF$ to define the \textit{Full-Processor Vulnerability Factor} ($FPVF$) for a given application workload. 
We define $FPVF$ as the ratio of the total number of vulnerable bits (\textit{VulnerableBits}) in the processor pipeline for the duration they are vulnerable (\textit{VulnerableTime}) to the total number of bits in the processor pipeline (\textit{TotalBits}) for the total duration of application execution (\textit{TotalTime}). 
It is computed using the following equation:
\begin{center}
$FPVF(T_q,PC_j) = \frac{\sum_{\forall C_{(j,k)}}VulnerableBits(C_{(j,k)}) \times VulnerableTime(C_{(j,k)})}{\sum_{\forall C_{(j,k)}}TotalBits(C_{(j,k)}) \times TotalTime(C_{(j,k)})}\times 100$
\end{center}

%% file: sections/section4.tex
\section{Heterogeneous Reliability Modes of Out-of-Order Superscalar Cores}
\label{sec:HHM}

\begin{figure*}[t]
	\centering
	\captionsetup{justification=raggedright,singlelinecheck=false}
	\includegraphics[width = \linewidth]{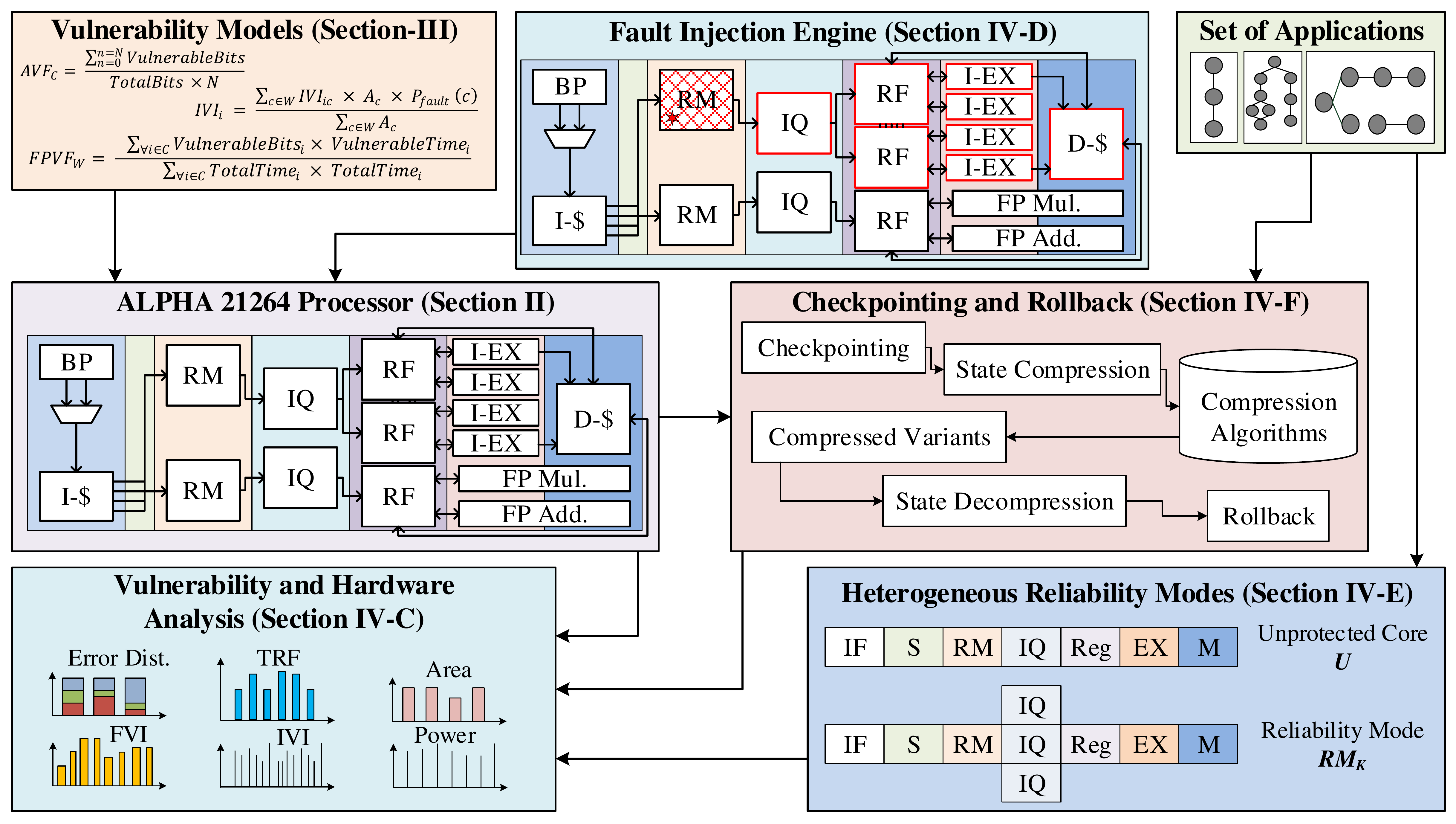}
	\caption{\textbf{Overview of Our Architecture-Space Generation and Exploration Methodology for Hardening Out-of-Order Superscalar Heterogeneous Multi-Core Processors.}}
	\label{fig:HHM}
\end{figure*}

\subsection{Methodology Overview}
\label{subsec:Methodology}
Fig.~\ref{fig:HHM} presents an overview of our methodology for designing and exploring heterogeneous reliability modes for out-of-order superscalar multi-core processors. 
Our methodology targets two approaches for designing heterogeneous reliability modes: (1) Redundancy, and (2) Checkpointing. 
To ensure reliable execution at the hardware layer, we propose hardening the processor's highly vulnerable pipeline components. 
These pipeline components are selected based on the initial fault-injection experiments, or on the AVF values that are estimated based on the number of vulnerable bits and vulnerable time of each component (see model description in Section~\ref{sec:SystemModel}). 
\changed{Furthermore, we ensure reliability by investigating state compression techniques that can reduce the size of checkpoint data.
Before moving on to our fault-injection and vulnerability analyses, we will present our experimental setup for better understanding.}


\subsection{Experimental Setup}
\label{subsec:ES}

To evaluate the vulnerability, power and area requirements of the proposed heterogeneous reliability modes, we have modified the well-established open-source tools like the cycle-accurate system simulator, gem5 \cite{binkert2011gem5} \changed{and HP's power and area estimator tool McPAT~\cite{McPATLi2009}.
Our extensions to these tool chains provide the following functionality}:
\begin{inlinelist}
    \item estimate the vulnerability of all pipeline components by determining their AVFs~\cite{1253181},
    \item support for heterogeneous reliability modes by hardening key pipeline components using component-level redundancy \cite{rehman2018hardware}, but not full-scale pipeline triplication all the time, and  
    \item checkpoint processor states using mechanisms like Distributed Multi-Threaded Checkpointing (DMTCP) \cite{5161063}\cite{DMTCP} and Hash-Based Incremental Checkpointing Tool (HBICT) \cite{Agarwal:2004:AIC:1006209.1006248}\cite{HBICT}.
\end{inlinelist}  
Due to its high customization capability, we use the Alpha 21264 four-issue out-of-order superscalar core \cite{755465} as our target platform.

\begin{figure}[h]
    \centering
    \captionsetup{justification=raggedright,singlelinecheck=false}
    \includegraphics[width = \linewidth]{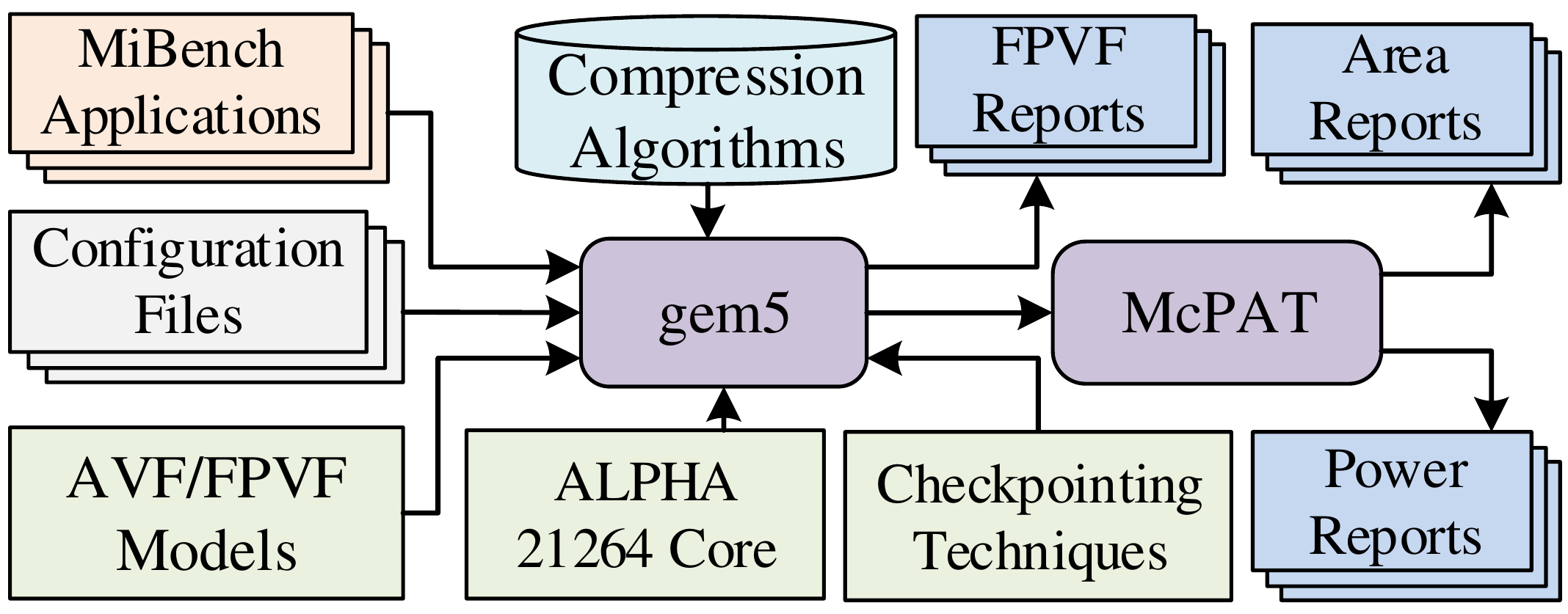}
    \caption{\textbf{Overview of our Experimental Setup.}}
    \label{fig:ES}
\end{figure}

Furthermore, we extend the concept of AVF towards the FPVF metric (see Section~\ref{sec:SystemModel}) to evaluate the impact of component hardening on the reliability mode, for a given application workload. 
To account for a wide range of applications, we evaluate the proposed heterogeneous reliability modes using the MiBench application benchmark suite. 
Fig.~\ref{fig:ES} presents an overview of our experimental setup.

\subsection{Vulnerability Analysis}
\label{subsec:VulAnalysis}

\begin{figure}[!t]
	\centering
	\captionsetup{justification=raggedright,singlelinecheck=false}
	\includegraphics[width = \linewidth]{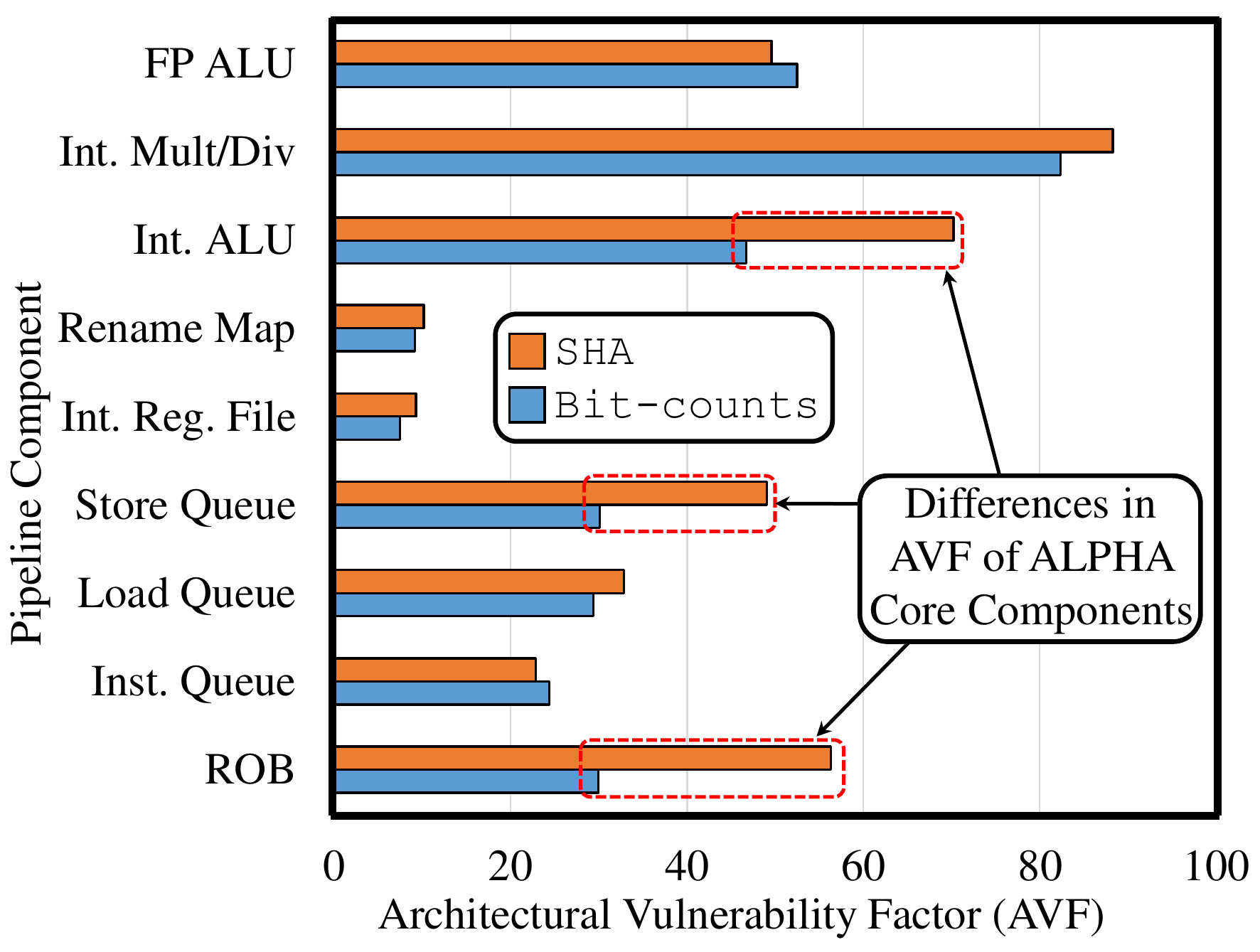}
	\caption{\textbf{Differences in AVF of Alpha 7 Pipeline Components under} \textbf{(}\texttt{SHA} \textbf{and} \texttt{Bit-counts}\textbf{ Workloads)}.}
	\label{fig:InitAnalysis}
\end{figure}

\begin{figure}[b]
	\centering
	\captionsetup{justification=raggedright,singlelinecheck=false}
	\includegraphics[width = \linewidth]{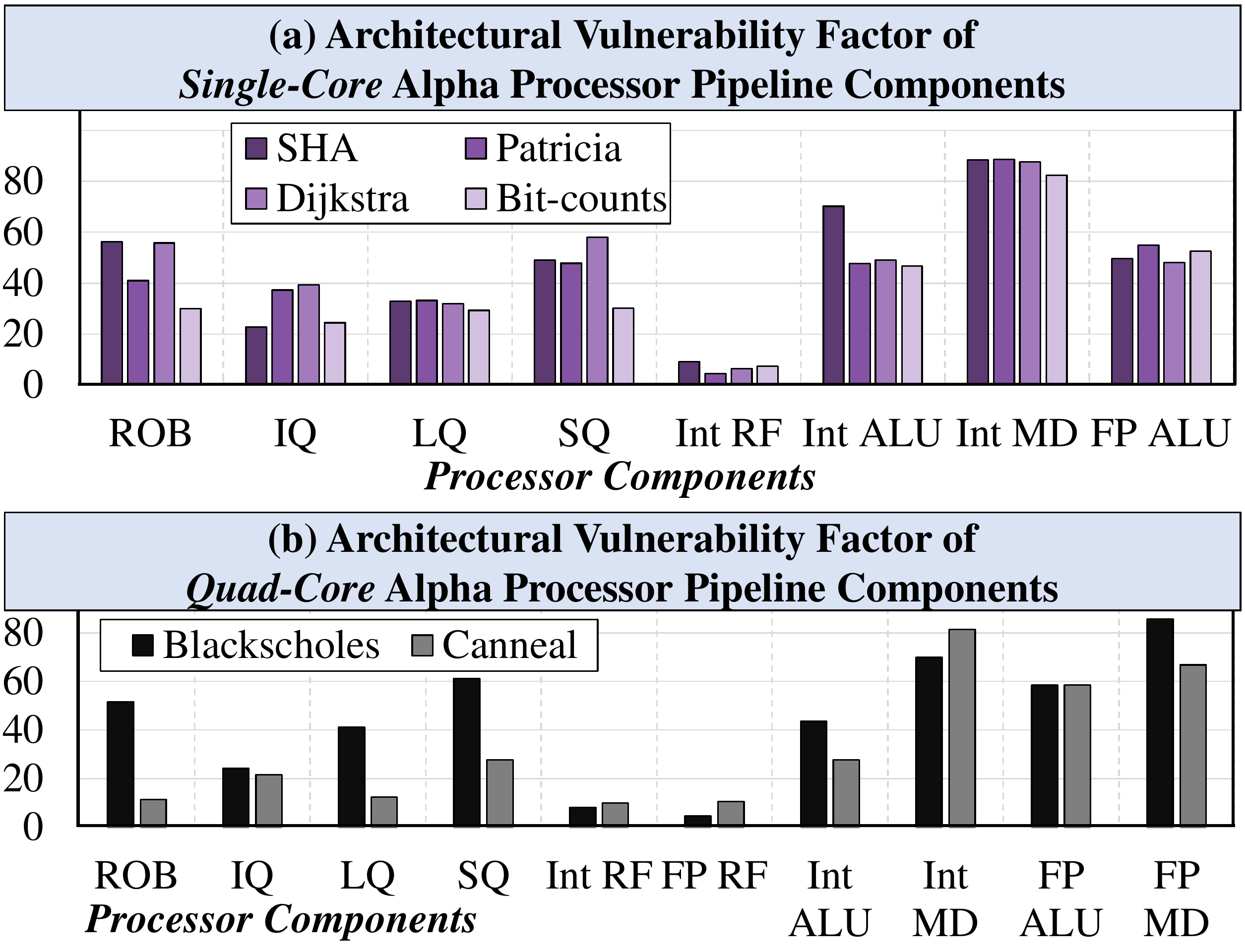}
	\caption{\textbf{AVF Distribution of Key Pipeline Components in Single- and Multi-Core Alpha 7 Processors.}}
	\label{fig:VulAnalysis2}
\end{figure}

We evaluate the vulnerability of an O3 superscalar Alpha 21264 core components \cite{755465} for the \texttt{Bit-counts}, \texttt{SHA}, \texttt{Dijkstra}, and \texttt{Patricia} application workloads~\cite{990739}.
We analyze the vulnerability of the following key pipeline components: 
\begin{itemize}
    \item Re-order Buffer (ROB),
    \item Issue Queue (IQ), 
    \item Load Queue (LQ), 
    \item Store Queue (SQ),
    \item Integer, Floating Pt. Register Files (RF),
    \item Rename Map (RM), 
    \item Integer ALU (Int. ALU),
    \item Floating Point ALU (FP ALU),
    \item Integer Multiply/Divide (Int. MD), and
    \item Floating Point Multiply/Divide (FP MD).
\end{itemize}
The results of our vulnerability analyses are presented in Figs.~\ref{fig:InitAnalysis} and \ref{fig:VulAnalysis2}.

From the results obtained, we make the following \textit{\textbf{key observations}}:
\begin{itemize}
    \item The AVFs of the different pipeline components vary for different application workloads. 
    \item We have identified three key pipeline components (Integer ALU, Store Queue, and Re-order Buffer) that are more vulnerable during the execution of \texttt{SHA}, when compared to \texttt{Bit-counts}.
    \item Similarly, the re-order buffer is ~$27\%$ and ~$46\%$ less vulnerable to soft errors during the execution of \texttt{Patricia} and \texttt{Bit-counts}, when compared to workloads like \texttt{SHA} and \texttt{Dijkstra}.
    \item Similar differences in component-AVFs can be observed when varying multi-threaded application workloads, from the PARSEC benchmark suite, are executed on a multi-core processor, as shown in Fig.~\ref{fig:VulAnalysis2}.
\end{itemize}

These components have different AVFs because of the type of instructions being executed and their application-specific properties (compute or memory-intensive, instruction-level parallelism, cache hit/miss rate, etc.). 
For example, components like the Re-order Buffer and the Store Queue are more vulnerable in \texttt{SHA} because of higher levels of instruction-level parallelism and more store instructions.

Based on this analysis, we can infer that hardening certain components of the pipeline increases the reliability of a core more than hardening the other components. 
Therefore, we generate a wide range of reliability-heterogeneous Alpha cores, and explore this architectural-space in terms of reliability, power, and area, to select a configuration that increases the reliability of application executions while decreasing the area/power overhead.

\subsection{Fault Injection}
\label{subsec:FI}

\setcounter{figure}{10}
\begin{figure*}[!b]
	\centering
	\captionsetup{justification=raggedright,singlelinecheck=false}
	\includegraphics[width = \linewidth]{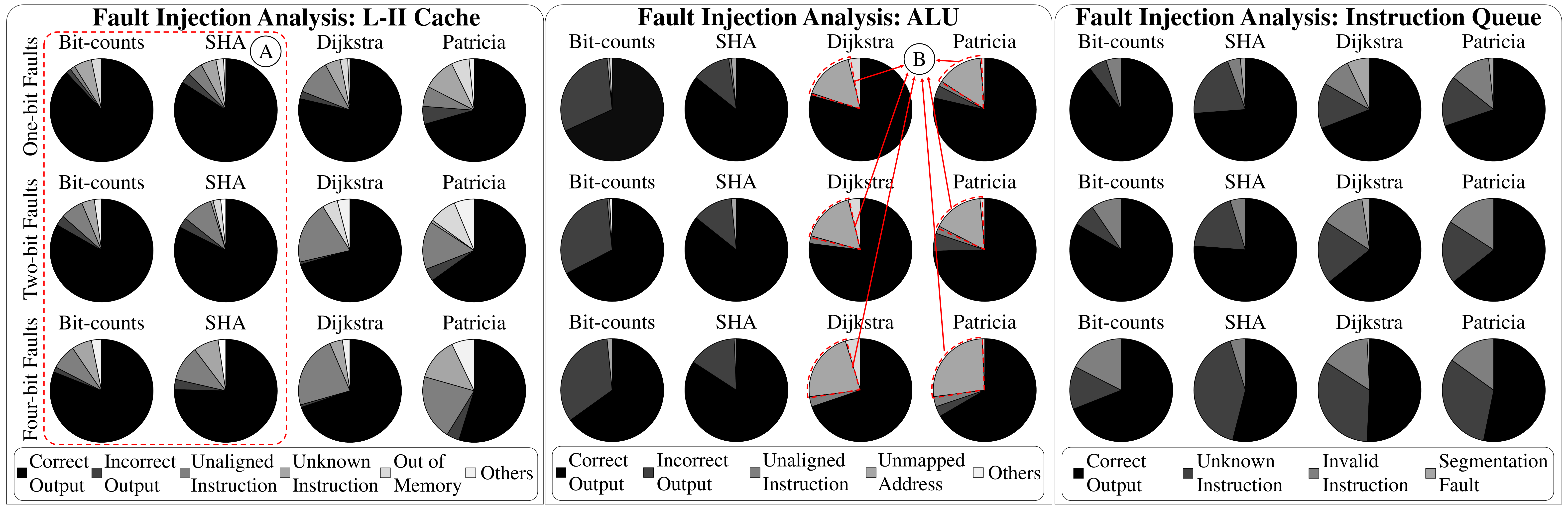}
	\caption{\textbf{Error Rate of Three Pipeline Components (L2 Cache, ALU, Instruction Queue) in the Alpha 7 Processor.}}
	\label{fig:FIE}
\end{figure*}

Fault injection techniques are typically used to study, analyze and evaluate the behaviour of a system susceptible to faults~\cite{SimulationGPU,ZiadeFaultInjection,RehmanCODES}.
The fault model for the ALPHA core components is based on single- and multi-bit transient faults. The soft error rate for each component is defined as the product of error rate and the component's AVF. 
The soft error rate of the processor's pipeline components have been derived from the works presented in~\cite{SoftErrorProblem}\cite{DixitSER}.
To account for a component's spatial vulnerability ($N_{FI}$), the number of faults injected in a pipeline component is proportional to its on-chip area.
We define $P_{flip}$ as the probability that a high-energy particle strike leads to a change in the logic state of a pipeline component.
Furthermore, to facilitate fast simulation, the faults are injected in the region of interest, the components, registers, and cache lines used by the application. 
The application output is classified into 3 major categories, namely,
\begin{inlinelist}
    \item correct output,
    \item incorrect output, and
    \item program failures ($N_{error}$), which comprise of multiple scenarios such as unaligned instruction, unmapped address, and segmentation fault.
\end{inlinelist}
The error rate ($P_{error}$) of a transient fault in the component leading to an error in the application execution is defined as follows:

\begin{center}
    $P_{error} = P_{flip} \times \frac{N_{error}}{N_{FI}}$
\end{center}

\setcounter{figure}{9}
\begin{figure}[!t]
	\centering
	\captionsetup{justification=raggedright,singlelinecheck=false}
	\includegraphics[width = \linewidth]{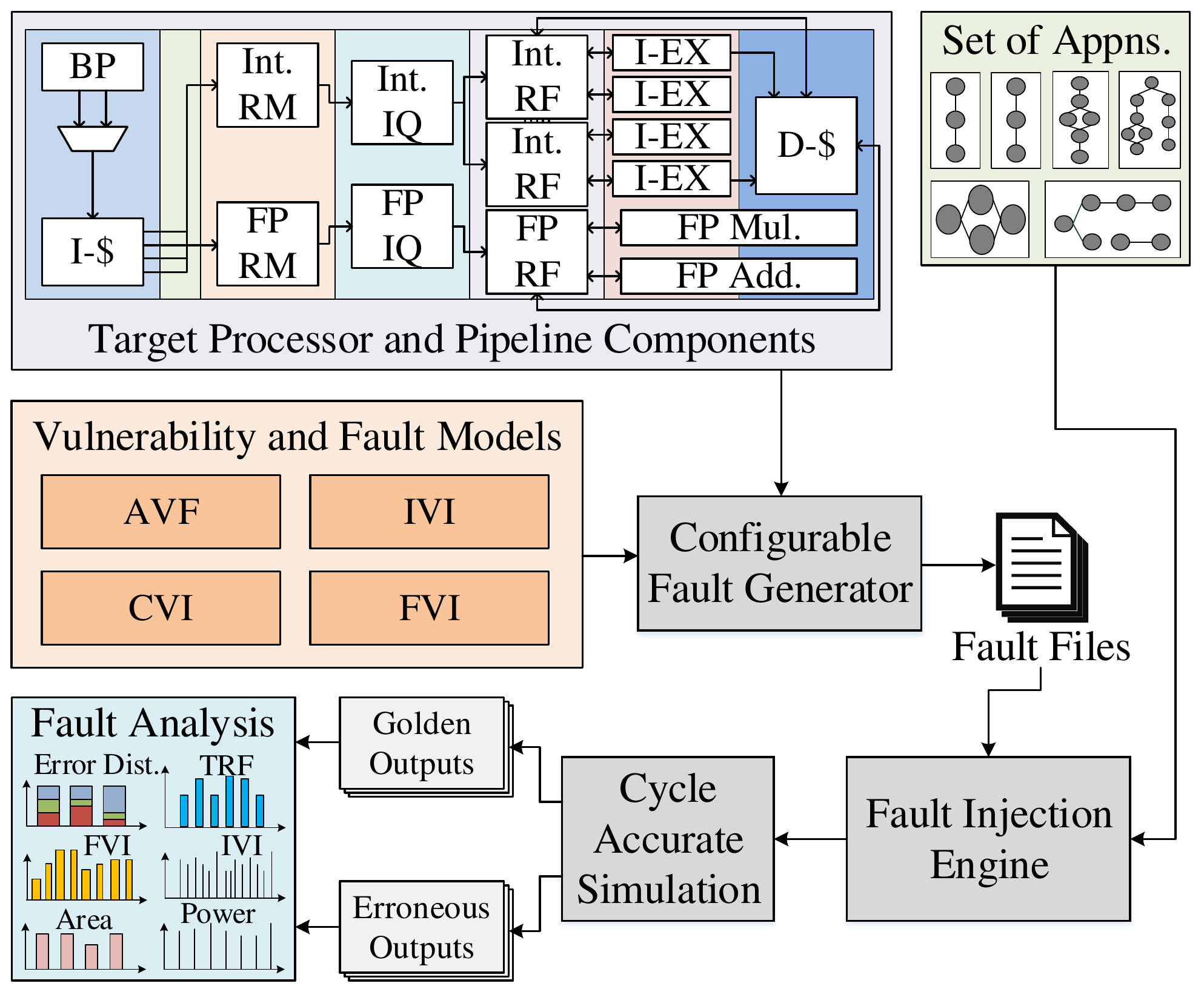}
	\caption{\textbf{Overview of the Fault Injection Methodology for Analyzing Processor Component Vulnerabilities.}}
	\label{fig:FIS}
\end{figure}

An overview of the methodology used to inject and analyze faults in various pipeline components is presented in Fig.~\ref{fig:FIS}.
Based on the vulnerability and fault models presented in Section~\ref{sec:SystemModel} and the configuration of the target processor, including its pipeline components, we generate a list of fault files, that is provided as an input to the fault injection engine.
This is used to insert faults/bit-flips into the target processor platform during the application's execution using a cycle-accurate simulator, i.e., gem5. 
The architectural parameters for the Alpha processor and the fault injection experiments are illustrated in tables~\ref{table3} and~\ref{table4}.
We study the output obtained from these simulations, which contains a list of correct and erroneous outputs.
These outputs are then compared against the golden execution to estimate the type of error and the frequency of these error occurrences for various pipeline components.
A subset of the results obtained from this experiment are illustrated in Fig.~\ref{fig:FIE}.

\begin{table}[h]
	\centering
	\caption{\protect\centering\textbf{Processor Parameters for Vulnerability Analyses Experiments}}
	\begin{tabular}{|r|l|}
		\cline{1-2}
		\multicolumn{1}{|r|}{\textbf{Parameter}} & \multicolumn{1}{|l|}{\textbf{Value}} \\ \hline
		\multicolumn{1}{|r|}{Core} & Alpha 21264 \\ \hline
		\multicolumn{1}{|r|}{Frequency} & \multicolumn{1}{|l|}{2 GHz} \\ \hline
		\multicolumn{1}{|r|}{Simulation Mode} & \multicolumn{1}{|l|}{Syscall Simulation}        \\ \hline
		\multicolumn{1}{|r|}{L1 Cache (I-\$ and D-\$)} & \multicolumn{1}{|l|}{32kB, 2-way, 64B, 2 cycles}        \\ \hline
		\multicolumn{1}{|r|}{L2 Cache} & \multicolumn{1}{|l|}{256kB, 2-way, 64B, 20 cycles}        \\ \hline
		\multicolumn{1}{|r|}{Cache Policy} & \multicolumn{1}{|l|}{Snooping Coherence, LRU}        \\ \hline
		\multicolumn{1}{|r|}{TLB} & \multicolumn{1}{|l|}{Data: 64, Instruction: 48}        \\ \hline
		\multicolumn{1}{|r|}{Re-order Buffer} & \multicolumn{1}{|l|}{192 Entries}        \\ \hline
		\multicolumn{1}{|r|}{Instruction Queue} & \multicolumn{1}{|l|}{64 Entries}        \\ \hline
		\multicolumn{1}{|r|}{Load-Store Queues} & \multicolumn{1}{|l|}{32 Entries}        \\ \hline
		\multicolumn{1}{|r|}{Register File} & \multicolumn{1}{|l|}{Integer: 256, Floating-Pt.: 256}        \\ \hline
		\end{tabular}
	\label{table3}
\end{table}

\begin{table}[h]
	\centering
	\caption{\protect\centering\textbf{Parameters for Fault Injection Experiments}}
	\begin{tabular}{|L{1.5cm}|C{4cm}|C{1.5cm}|}
		\cline{1-3}
		\textbf{Parameter} & \textbf{Description} & \textbf{Properties/ Values}\\ \hline
		
		Distribution & Distribution models \newline for fault generation & Random\\ \hline
		
		Bit Flips & Min/Max number of bits flipped & 1/1, 1/2, ...\\ \hline
		
		Fault Probability & Probability that \newline strike becomes a fault~\cite{baumann2005radiation} & 10\%-100\% \\ \hline
		
		Fault Location & List of target \newline processor components & Register file, PC, IQ, etc.\\ \hline
		
		Processor Layout/ Area & Size of the complete target device & Gate-equivalents or mm$^2$~\cite{GaislerDSN}\\ \hline
		
		Component Area & Area of different processor components given as \newline percentage of processor area & 0\%-100\%\\ \hline
		
		Place and Altitude & City and altitude at which the device is used to \newline determine the flux rate (N$_{Flux}$) & Oslo, 1-20km\\ \hline
		
		Frequency & Operating frequency \newline of the processor & 50, 100 MHz\\ \hline
		\end{tabular}
	\label{table4}
\end{table}

\setcounter{figure}{11}
\begin{figure*}[t]
	\centering
	\includegraphics[width = 0.934\linewidth]{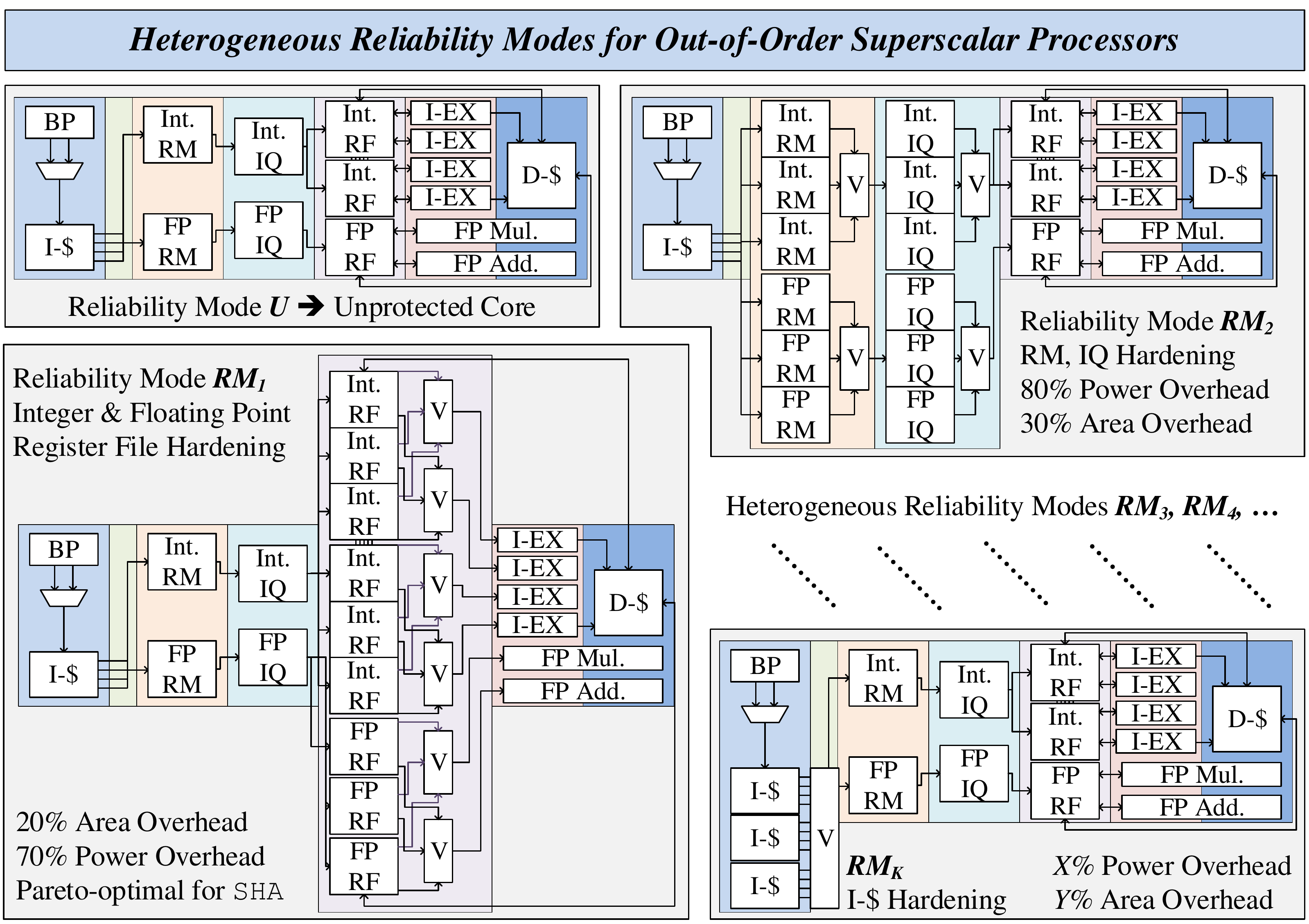}
	\caption{\textbf{Heterogeneous Reliability Modes for ALPHA 7 Out-of-Order Superscalar Processor.}}
	\label{fig:HM}
\end{figure*}

The results in Fig.~\ref{fig:FIE} depict the error rate of three pipeline components, namely, Level-II Cache, Integer Arithmetic Logic Unit, and Instruction Queue.
Faults injected in the L2-cache lead to four major types of error and correct output. The rest of the types are classified into the ``others'' category. The four major error categories are:
\begin{inlinelist}
    \item incorrect output,
    \item unaligned instruction,
    \item unknown instruction, and
    \item out of memory.
\end{inlinelist}
The label \textit{A} depicts the applications with a higher percentage of correct output, when compared to the others.
On average, the \texttt{Bit-counts} and \texttt{SHA} applications produce a correct output more than 80\% of the time, whereas \texttt{Dijkstra} and \texttt{Patricia}, on average produce a correct output less than ~70\% and ~60\% of the time.
This is due to the lower number of load and store instructions in the two applications present in label \textit{A}, when compared to the other two.
Therefore, the probability of a soft error in L2-cache leading to an error during the execution is higher in an application with a relatively higher number of load and store instructions as compared to the others.
Similarly, the label \textit{B} depicts the percentage of fault injection experiments that lead to an unmapped address.
As explained in the earlier example, due to the higher number of load and store instructions in \texttt{Dijkstra} and \texttt{Patricia}, the large number of unmapped addresses can be attributed to the corruption of bits during address generation.
Similarly, due to their compute-intensive nature, a higher number of incorrect outputs are generated by faults injected in an ALU during the execution of applications like \texttt{bit-counts} and \texttt{SHA}.
Faults injected in the Instruction Queue cause three major types of error, namely, 
\begin{inlinelist}
    \item unknown instruction,
    \item invalid instruction, and
    \item segmentation fault.
\end{inlinelist}

\subsection{Heterogeneous Reliability Modes for ALPHA Cores}
\label{subsec:HRM}

\begin{figure*}[t]
	\centering
	\captionsetup{justification=raggedright,singlelinecheck=false}
	\includegraphics[width = \linewidth]{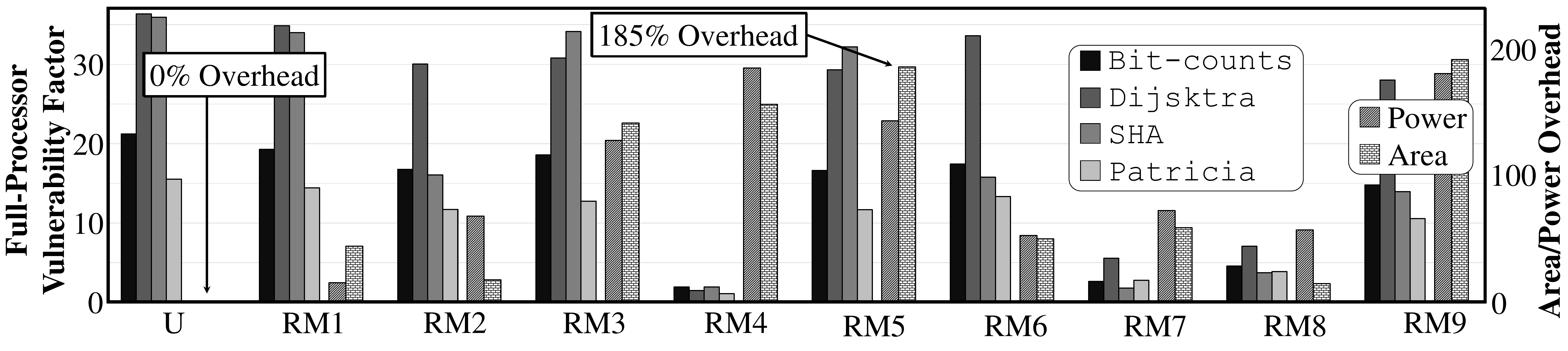}
	\caption{\textbf{Full-Processor Vulnerability Factor (FPVF) and Power/Area Trade-off of Our Heterogeneous Reliability Modes for Different MiBench Applications.}}
	\label{fig:FPVF}
\end{figure*}

\begin{figure*}[h]
	\centering
	\captionsetup{justification=raggedright,singlelinecheck=false}
	\includegraphics[width = \linewidth]{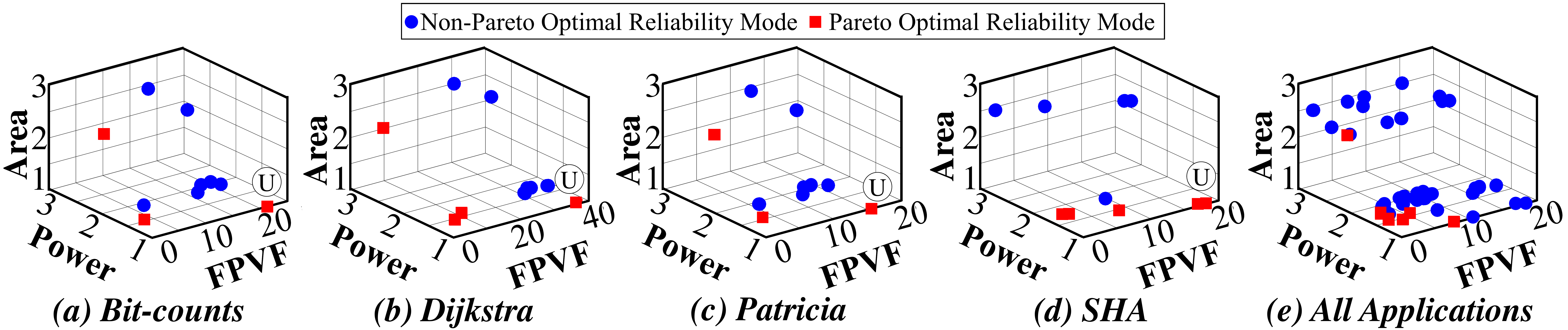}
	\caption{\textbf{Design Space Exploration of Our Heterogeneous Reliability Modes for MiBench Applications.}}
	\label{fig:DSE}
\end{figure*}

As discussed in Section~\ref{subsec:VulAnalysis}, the AVF of the pipeline components varies for the different application workloads. 
Hence, we propose to harden a combination of the key pipeline components \changed{in out-of-order superscalar processors}, instead of employing full-scale TMR across the complete pipeline, to increase core reliability while reducing the area and power overheads of full-scale TMR. 
This generates a design space of multiple heterogeneous reliability modes (RM), nine of which are illustrated in this work (and unprotected core). 
Table~\ref{table1} presents our list of nine proposed heterogeneous RM and the components that are hardened in these modes using TMR.
Hardened components have three instances with the same inputs, and a voter circuit at the output to determine the majority.
An overview of the proposed heterogeneous reliability modes for Alpha 7 processor is presented in Fig.~\ref{fig:HM}.

We evaluate the vulnerability of our heterogeneous reliability modes by executing applications from the MiBench application benchmark to estimate the FPVF for each scenario. 
We also evaluate the area and power overheads incurred by each reliability mode. 
These results are illustrated in Fig.~\ref{fig:FPVF}. 

From the results obtained, we make the following \textit{\textbf{key observations}}:
\begin{itemize}
    \item Different heterogeneous reliability modes can reduce the full-processor vulnerability to different extents depending upon the properties of the executing application. 
    For example, reliability modes like RM2, RM6, and RM9 reduce the processor vulnerability of \texttt{SHA} by more than 50\%, but not of \texttt{Dijkstra}, even though they have similar vulnerabilities in all other reliability modes. 
    \item Hardening specific components in the pipeline can significantly reduce the overall processor vulnerability. 
    For example, key components like Rename Map (RM) and Reorder Buffer (ROB) effectively reduce the FPVF for all applications, as shown by the heterogeneous reliability modes RM4, RM7 and RM8. 
    However, utilizing these hardening modes incurs significant area and power overheads. 
    \item \changed{Certain heterogeneous reliability modes are very effective in reducing the FPVF by a large margin for very small area/power overhead.
    For example, RM2 and RM6 reduce the FPVF by more than $50$\% for <$75$\% area and power overheads when executing \texttt{SHA}.
    \item Hardening all pipeline components without hardening the most highly vulnerable component of the system introduces very high overheads without reducing the vulnerability of the system significantly.
    This is illustrated by the reliability mode RM9, in which the ROB is not hardened. This reliability mode has area and power overheads close to \textasciitilde$200$\% with insignificant reductions in FPVF, when compared to RM4, which significantly reduces the FPVF for comparatively lower overheads.
    }
\end{itemize}

\begin{table}[h]
	\centering
	\caption{\protect\centering\textbf{Proposed Heterogeneous Reliability Modes}}
	\begin{tabular}{|c|l|}
		\cline{1-2}
		\multicolumn{1}{|c|}{\textbf{Reliability Mode}} & \textbf{Components Hardened} \\ \hline
		\multicolumn{1}{|c|}{U} & \multicolumn{1}{|l|}{Unprotected} \\ \hline
		\multicolumn{1}{|c|}{RM1} & \multicolumn{1}{|l|}{RF}        \\ \hline
		\multicolumn{1}{|c|}{RM2} & \multicolumn{1}{|l|}{IQ, RM}        \\ \hline
		\multicolumn{1}{|c|}{RM3} & \multicolumn{1}{|l|}{IQ, LQ, SQ}        \\ \hline
		\multicolumn{1}{|c|}{RM4} & \multicolumn{1}{|l|}{IQ, LQ, SQ, RM, ROB}        \\ \hline
		\multicolumn{1}{|c|}{RM5} & \multicolumn{1}{|l|}{RF, IQ, LQ, SQ}        \\ \hline
		\multicolumn{1}{|c|}{RM6} & \multicolumn{1}{|l|}{RF, RM}        \\ \hline
		\multicolumn{1}{|c|}{RM7} & \multicolumn{1}{|l|}{RF, RM, ROB}        \\ \hline
		\multicolumn{1}{|c|}{RM8} & \multicolumn{1}{|l|}{RM, ROB}        \\ \hline
		\multicolumn{1}{|c|}{RM9} & \multicolumn{1}{|l|}{RF, IQ, LQ, SQ, RM}        \\ \hline
		\end{tabular}
	\label{table1}
\end{table}

\begin{table}[h]
	\centering
	\caption{\protect\centering\textbf{Pareto-Optimal Reliability Modes\newline for MiBench Applications}}
	\begin{tabular}{|c|l|}
		\cline{1-2}
		\multicolumn{1}{|c|}{\textbf{Application}} & \textbf{Pareto-Optimal Reliability Modes} \\ \hline
		\multicolumn{1}{|l|}{\texttt{Bit-counts}} & \multicolumn{1}{|l|}{U, RM4, RM7} \\ \hline
		\multicolumn{1}{|l|}{\texttt{Dijkstra}} & \multicolumn{1}{|l|}{U, RM4, RM7, RM8}        \\ \hline
		\multicolumn{1}{|l|}{\texttt{Patricia}} & \multicolumn{1}{|l|}{U, RM4, RM7}        \\ \hline
		\multicolumn{1}{|l|}{\texttt{SHA}} & \multicolumn{1}{|l|}{U, RM2, RM6, RM7, RM8}        \\ \hline
		\multicolumn{1}{|l|}{\texttt{All}} & \multicolumn{1}{|l|}{U, MR4, RM7, RM8}        \\ \hline
		\end{tabular}
	\label{table2}
\end{table}

\begin{algorithm}[!t]
	\caption{Pareto-Frontier Extraction}
	\label{Algo1}
	\begin{algorithmic}[1]
		\Input $\{FPVF, A, P\} \forall RM_{\forall i \in [1,K]}$
		\Output $OptimalReliabilityModes~(ORM)$
		\State $TempSignal = 0;$
		\State $TempArray(3,K) = 0;$
		\State $TempArray2(3,K) = 0;$
		\State $B = [FPVF, Area, Power];$
		\For{$k$ $\gets$ $1$ $to$ $3$}
		\State $j=0;$
		\State $temp=B(k,:);$
		\For{$i$ $\gets$ $1$ $to$ $3$}
		\If{$i!=k$}
		\State $j=j+1;$
		\State $TempArray2(j,:)=temp-B(i,:);$
		\EndIf
		\EndFor
		\If{$TempArray2(1:j,:)<0$}
		\State $TempSignal = TempSignal+1;$
		\State $TempArray(TempSignal,:) = temp;$
		\EndIf
		\EndFor
		\If{$TempSignal>=1$}
		\State $ORM=TempArray(1:TempSignal,:);$
		\EndIf
	\end{algorithmic}
\end{algorithm}

Using the data gathered from the simulation of our designs, we perform a design space exploration that trades-off FPVF, area, and power overheads to extract the pareto-optimal designs that suit the target application best. 
The pseudo code of the pareto-frontier extraction algorithm is presented in Algorithm~\ref{Algo1}.
The corresponding results are illustrated in Fig.~\ref{fig:DSE}. 
The \textit{x}-axis denotes the FPVF, whereas the \textit{y}- and \textit{z}-axes denote the power and area overheads, respectively. 
The design labeled \textbf{\textit{U}} in all applications is the unprotected core that is highly vulnerable to soft errors. 
As it does not deploy any redundancy measures, it has zero area and power overhead, and hence lies on the pareto-front. 
The pareto-optimal reliability modes for the applications are presented in Table~\ref{table2}. 
RM4 is pareto-optimal for all applications except \texttt{SHA}. 
The register file is highly vulnerable to soft errors during the execution of \texttt{SHA} and needs to be hardened to reduce its vulnerability. 
The reliability mode \textit{\textbf{RM7} is pareto-optimal for all four applications} and reduces the FPVF on average by 87\% with average area and power overheads of 10\% and 43\%, respectively.

\changed{A super-set of the pareto-optimal reliability modes for all these applications can be selected to design a heterogeneous multi-core processor.
We can build the chip by selecting the reliability modes from this super-set such that the form-factor and cost constraints are adhered to.
At run-time, the required reliability modes can be switched-on/-off depending upon the power constraints of the system.}

\subsection{Run-Time System}
\label{subsec:RTS}

\begin{figure}[t]
	\centering
	\captionsetup{justification=raggedright,singlelinecheck=false}
	\includegraphics[width = \linewidth]{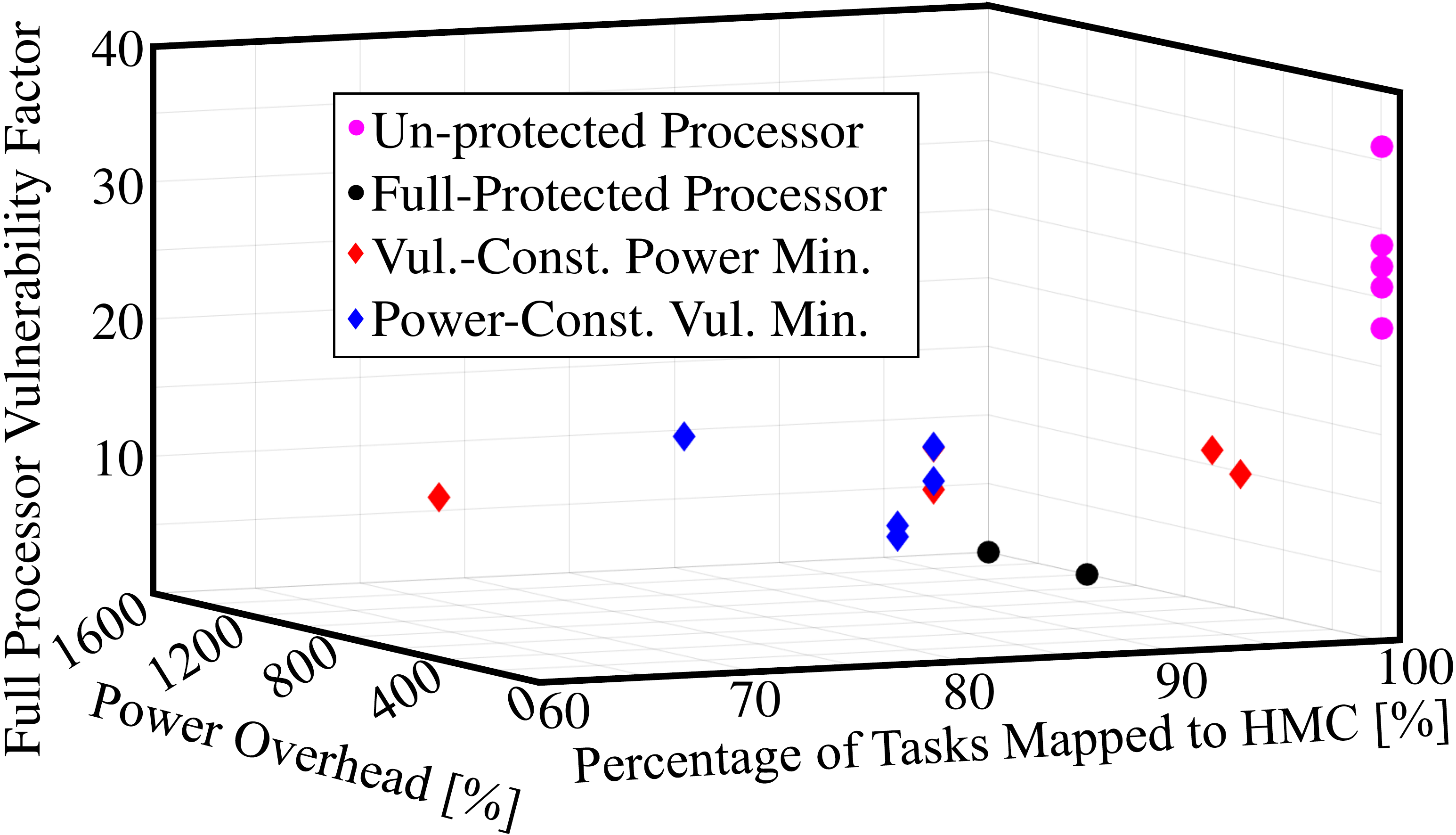}
	\caption{\textbf{Run-Time Task Mapping Analysis of HMC.}}
	\label{fig:RTS}
\end{figure}

Although this work focuses mostly on the design-time aspects of achieving heterogeneous reliability in out-of-order superscalar processors, in this sub-section we present a brief overview of a run-time system analysis using our proposed heterogeneous multi-core processor.
Our $HMC$ is a 10-core processor that is composed of all the 10 heterogeneous reliability modes discussed in sub-section~\ref{subsec:HRM}. 
We illustrate the benefits of our reliability modes by executing 5 application workload mixes, the compositions of which are presented in Table~\ref{table6}, on the 10-core heterogeneous processor to evaluate the power-overheads and FPVF of the $HMC$ for each workload mix.
The task-to-core mapping can be done using one of the following heuristics:
\begin{figure}[t]
	\centering
	\captionsetup{justification=raggedright,singlelinecheck=false}
	\includegraphics[width = \linewidth]{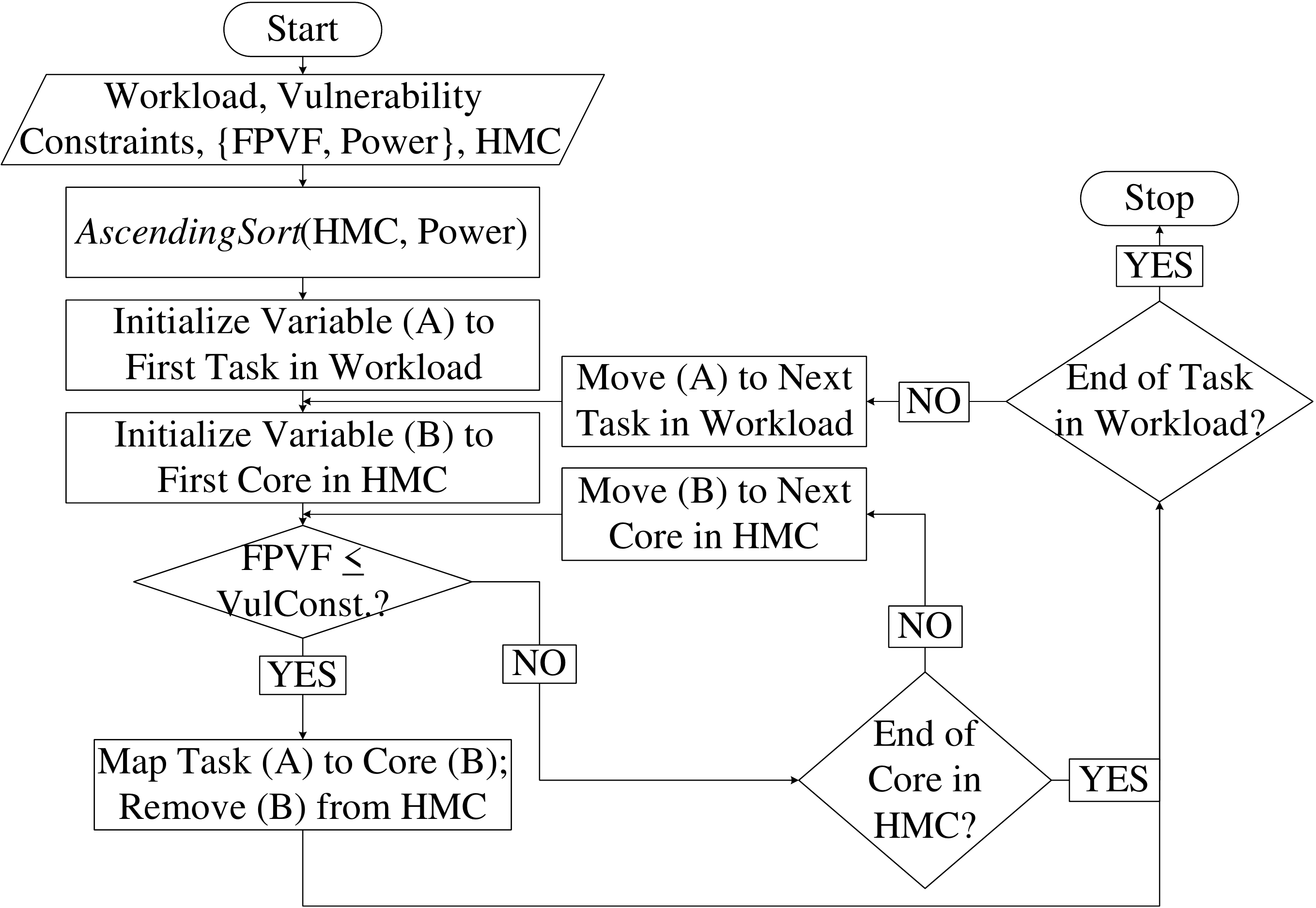}
	\caption{\textbf{Flowchart Illustrating the Vulnerability-Constrained Power Minimization Task-to-Core Mapping Policy.}}
	\label{fig:TMA1}
\end{figure}
\begin{figure}[t]
	\centering
	\captionsetup{justification=raggedright,singlelinecheck=false}
	\includegraphics[width = \linewidth]{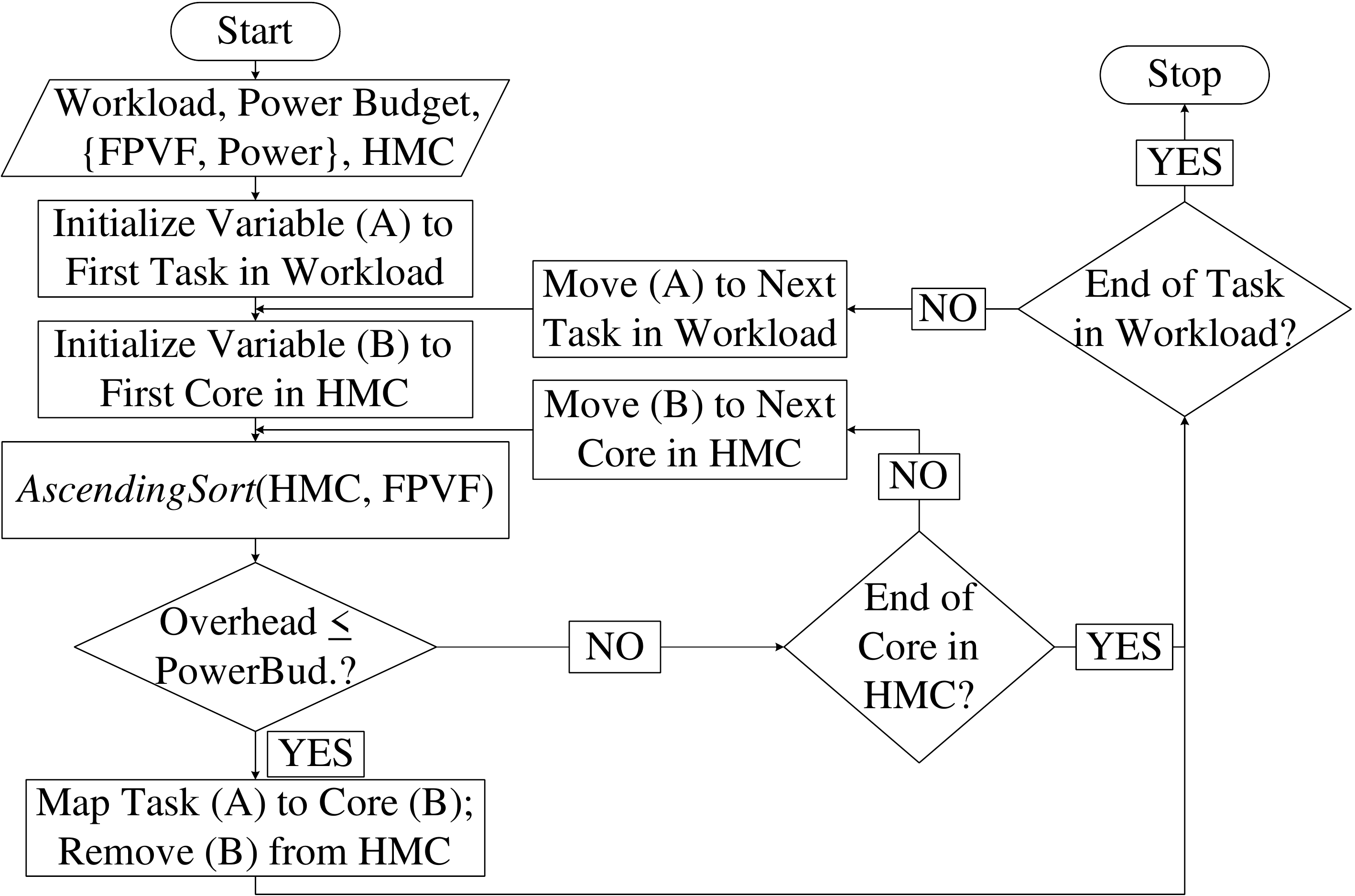}
	\caption{\textbf{Flowchart Illustrating the Power-Constrained Vulnerability Minimization Task-to-Core Mapping Policy.}}
	\label{fig:TMA2}
\end{figure}
\begin{enumerate}[leftmargin=*,label=(\arabic*)]
    \item \textbf{Vulnerability-Constrained Power Minimization:} In this technique (see Fig.~\ref{fig:TMA1}), we impose a vulnerability constraint on each task in the mix, i.e., each task is only mapped sequentially to a core that can successfully execute the task under the imposed vulnerability constraint.
    If a convenient core (one that satisfies vulnerability constraint) is not available, then the task is not scheduled immediately.
    The goal of this approach is to minimize the power overhead of the complete processor.
    \item \textbf{Power-Constrained Vulnerability Minimization:} This approach (Fig.~\ref{fig:TMA2}) imposes a constraint on the maximum power overhead of the whole processor, i.e., the task-to-core mapping is stalled when the power constraint is exceeded, which is an overhead of 100\% for each task in the mix.
    The goal of this task mapping policy is to minimize the FPVF.
\end{enumerate}
The results of this evaluation are presented in Fig.~\ref{fig:RTS}, in which we make the following \textit{\textbf{key observations}}:
\begin{itemize}
    \item The proposed reliability modes can be deployed in a heterogeneous multi-core processor to reduce the power overheads of the executing the application workloads, based on the application's workload requirement.
    \item The proposed reliability modes can either be used to minimize the power overhead or the full-processor vulnerability factor as illustrated by the two task mapping policies.
\end{itemize}
Although $100$\% task mapping is not achieved as in the un-protected or full-protected case, this can be resolved by efficiently selecting the reliability modes to be deployed in the $HMC$ considering the potential application workloads and/or by using a task mapping algorithm that can efficiently schedule the tasks to processor cores.

\begin{table}[h]
	\centering
	\caption{\protect\centering\textbf{Workload Mixes and their Application Compositions}}
	\begin{tabular}{|C{1.5cm}|C{5.5cm}|}
		\cline{1-2}
	    \textbf{Application Mix} & \textbf{Composition} \\ \hline
	    MIX-1 & [Bit-counts, Dijkstra, SHA, Patricia, \newline Bit-counts, Dijkstra, SHA, Patricia] \\ \hline
	    MIX-2 & [Bit-counts, Bit-counts, Bit-counts, Bit-counts, \newline Dijkstra, Dijkstra, Dijkstra, Dijkstra] \\ \hline
	    MIX-3 & [Bit-counts, SHA, Patricia, Bit-counts, \newline SHA, Patricia, Bit-counts, Patricia] \\ \hline
	    MIX-4 & [SHA, Patricia, SHA, Patricia, SHA, Patricia] \\ \hline
	    MIX-5 & [SHA, SHA, SHA, SHA, SHA, Dijkstra] \\ \hline
		\end{tabular}
	\label{table6}
\end{table}

\subsection{State Compression Techniques}
\label{subsec:SC}

\begin{figure}[!t]
	\centering
	\captionsetup{justification=raggedright,singlelinecheck=false}
	\includegraphics[width = 0.9\linewidth]{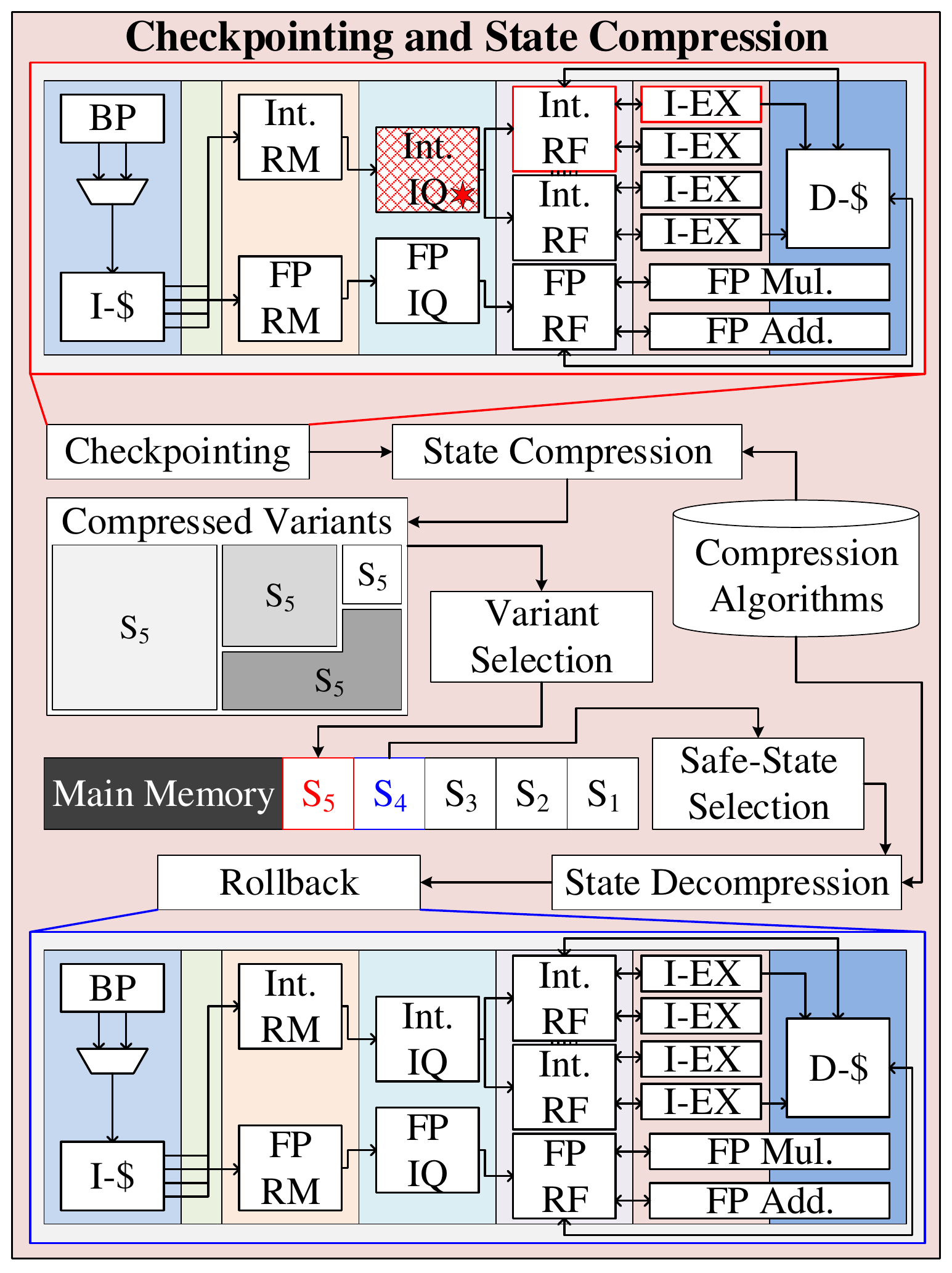}
	\caption{\textbf{Overview of the Methodology for Checkpointing and State Compression.}}
	\label{fig:SCMethodology}
\end{figure}

Checkpointing and Rollback is an effective way of guaranteeing reliability at the software layer by means of providing both spatial and temporal redundancy. 
A checkpoint is a snapshot of the processor state at any instant in time. 
Checkpoints allow the system to rollback to the previous safe states in case a failure is detected and re-execute instructions.

Fig.~\ref{fig:SCMethodology} presents an overview of the methodology that we use for checkpointing and state compression.
Checkpoints are typically inserted intermittently into the target application for periodic state retention and, if required, rollback to an earlier processor state, i.e., in case of faulty execution.
Typically, the collected processor's state information is stored in 
the main memory or off-chip nonvolatile memory, which can still be used for rollback in case of power-off.
In our case, to reduce the size of checkpointing data, we introduce another stage of state compression, that utilizes state-of-the-art compression techniques to generate a wide-range of compressed checkpoint variants.
The optimal compressed variant can be selected based on the system's resource constraints and available on-/off-chip memory.
In case a fault is detected in the current processor state, during the application execution, the previous safe-state is decompressed and rolled back to ensure the correct execution of the application.

The checkpointing mechanism deployed by gem5 comes with certain caveats. 
This technique does not preserve cache and pipeline states in a checkpoint because of which frequent restoration from such checkpoints results in performance loss, if deployed in real-world systems. 
Therefore, we explore techniques like DMTCP \cite{5161063}\cite{DMTCP} that checkpoints the Linux process. 
The back-end checkpointing mechanism of DMTCP is accessible to the programmer via numerous APIs. 
These APIs can be used in conjunction with the front-end gem5 pseudo-instructions for checkpoint creation/recovery. 
Since these software-based checkpoints are often large, the checkpoint is compressed using gzip and HBICT to save memory. 
HBICT \cite{Agarwal:2004:AIC:1006209.1006248}\cite{HBICT} provides DMTCP support for delta-compression (relative to the previous compression) which is further compressed using gzip (combination of lossless data compression algorithms like LZ77 and Huffman coding). 

\begin{figure}[h]
	\centering
	\captionsetup{justification=raggedright,singlelinecheck=false}
	\includegraphics[width = 0.85\linewidth]{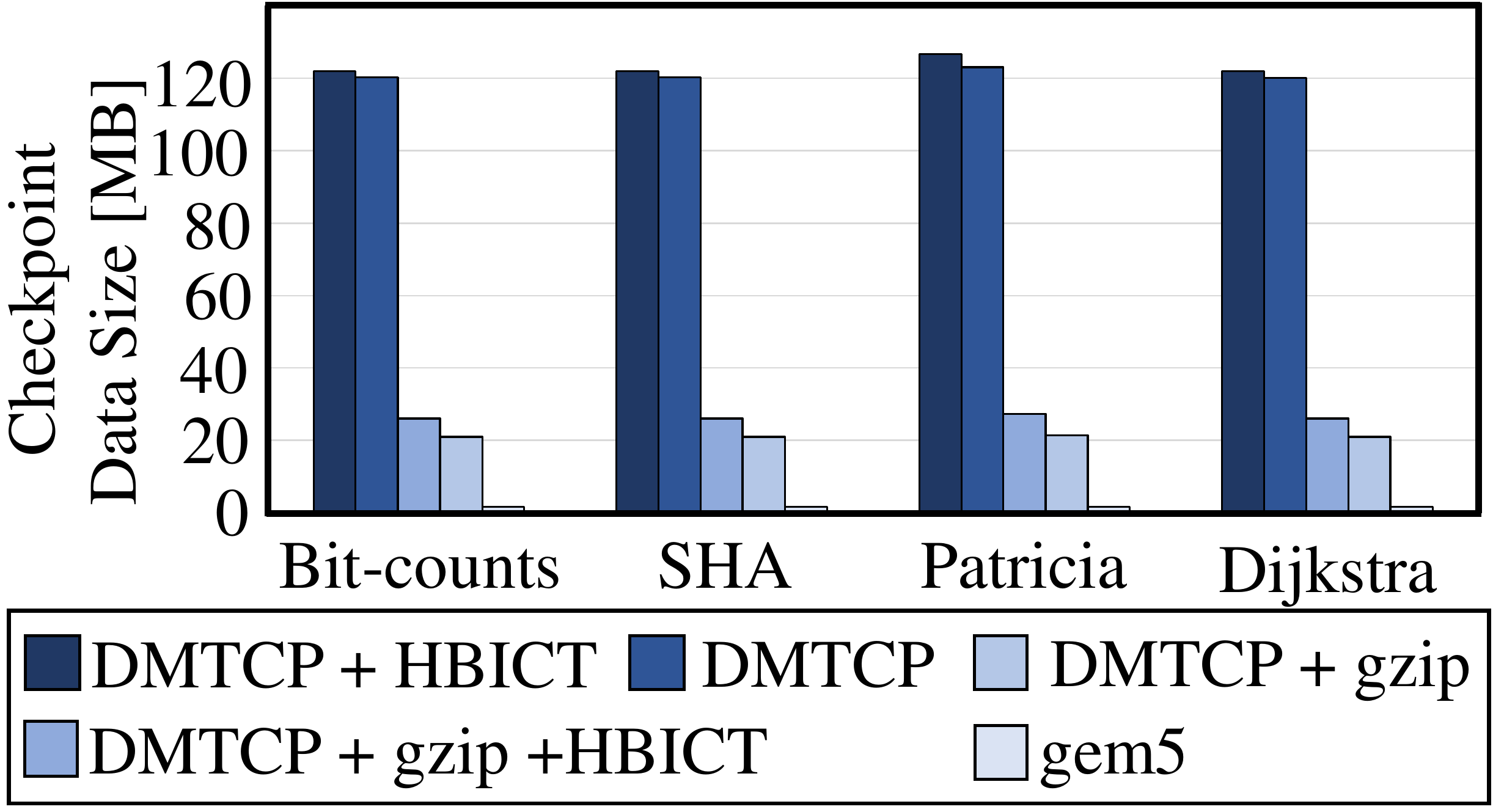}
	\caption{\textbf{Effectiveness of State Compression\hspace{\textwidth} Techniques in Reducing State Size.}}
	\label{fig:SC}
\end{figure}

We investigate the effectiveness of these techniques in all possible combinations, by applying them one after the other, on applications from the MiBench application benchmark suite by simulating them on the ALPHA core using gem5. 
The results of this experiment are presented in Fig.~\ref{fig:SC}.
From these results, we make the following \textit{\textbf{key observations:}}
\begin{itemize}
    \item the combination of DMTCP and gzip is highly successful in reducing the checkpoint size by \textasciitilde 6$\times$
    \item the combination of DMTCP, HBICT, and gzip techniques reduces the checkpoint size by \textasciitilde 5.7$\times$. 
\end{itemize}
HBICT, which utilizes delta-compression, requires all previous checkpoints for efficient rollback.
Since the base file size of HBICT+DMTCP is $1.03\times$ larger than the file size of DMTCP, the effectiveness of the combined state compression technique (DMTCP+HBICT+gzip), with respect to DMTCP, is reduced.

%% file: sections/section5.tex
\section{Related Work}
\label{sec:RW}
Reliability is a major research challenge that is being tackled by the community at large via global initiatives like the NSF's Variability Expedition\footnote{http://www.variability.org/} and DFG's SPP 1500 Priority Program\footnote{http://spp1500.itec.kit.edu}.
Research works from the academia and industry alike have addressed the challenges associated with technology scaling across the layers of the computing stack.

\changed{\textbf{Mitigation Strategies:}} The work in \cite{ErnstRazor} presents the \textit{Razor} approach, which can be used to dynamically detect and correct timing errors by monitoring the error rate at run-time to tune the circuit's supply voltage.
The adaptive approach presented in \cite{VadlamaniDATE2010} enables per-core dual modular redundancy (DMR) through the means of DVFS to offer a stable soft error rate (SER).
An OS-level dynamic reliability management system for heterogeneous architectures for achieving an optimal trade-off between reliability (lifetime) and power/performance efficiency is presented in \cite{baldassari2017dynamic}.
A software-level technique is presented in \cite{NahmsukTR}, which is used to detect errors by duplicating instructions during compile time by using different variables and registers for new instructions.
A software-controlled fault-tolerance scheme is proposed in \cite{ReisTACO} that allows programmers and designers to trade-off between performance and reliability based on the system's requirement.
Luo \textit{et al.} \cite{luo2014characterizing} quantify the tolerance of application to memory errors to propose several new hardware/software heterogeneous-reliability memory systems to reduce their vulnerabilities and data-center costs.

\changed{\textbf{Reliability Modeling:}} The work in \cite{SridharanHPCA} demonstrates the concept of Program Vulnerability Factor, which captures the architecture-level fault masking properties of the underlying program while exhibiting workload-driven changes in the AVF for all architectural components.
Li \textit{et al.} \cite{LinLiISLPED} analyze the correlation between soft-error rate and the energy consumption behaviour of on-chip data caches. This involves analyzing (1) the leakage energy optimizations on soft errors, and (2) the energy overheads of protecting on-chip memories against soft errors.
A software-level technique proposed in \cite{ShyeDSN07} introduces transient fault tolerance in a multi-core system by exploiting process-level redundancy (PLR) to create multiple application threads and compare them to ensure correct execution of the application.
A software-level approach to enable self-adaptive reliability for multi-/many-core systems is proposed in \cite{BolchiniJET} by activating redundancy measures based on the application's dependability requirements.
A simultaneous and redundantly threaded (SRT) processor is presented in \cite{ReinhardtISCA}, which provides transient fault tolerance with significantly higher performance. Redundant copies of the program threads are executed simultaneously on the SRT to ensure accurate application execution.
Kriebel \textit{et al.} \cite{kriebel2016reliability} analyze and present the reliability issues of on-chip memory systems to propose a reliability-aware reconfigurable last-level cache architecture that adapts the cache parameters to concurrently execute multi-threaded workloads at run-time in order to minimize their vulnerabilities.
A soft error-aware cache architectural space-exploration methodology is presented in \cite{ArunDATE2017} for varying the application workloads and cache parameters for the complete cache hierarchy.
An adaptive soft-error resilience (ASER) approach is presented in \cite{Kriebel:2014:AAS:2593069.2593094} by proposing and managing reliability-heterogeneous dark silicon many-core processors (darkRHPs). 
The proposed darkRHPs deploy redundancy at the architecture level, i.e., hardening either the full-processor pipeline of an in-order LEON3 processor and/or caches.
The work in \cite{kriebel2016variability} presents an approach that exploits the on-chip dark-silicon to synergistically mitigate reliability and variability challenges associated with transistor technology scaling.
\changed{An overview of different heterogeneous fault-tolerance schemes for both hardware and software layers is presented in \cite{rehman2018hardware}, which also provides an initial proof-of-concept of this work.}

This work, on the other hand, focuses on generating and exploring a wide-range of heterogeneous reliability modes using two key approaches, i.e., (1)~Redundancy, by hardening different combinations of the pipeline components for an out-of-order superscalar processor, and (2)~Checkpointing, by reducing the size of the checkpoint data using efficient compression techniques.

%% file: sections/section6.tex
\section{Conclusion}
\label{sec:conclusion}

In this work, we presented a novel design space generation and exploration methodology that is used to develop a wide range of heterogeneous reliability modes for out-of-order superscalar processors.
By analyzing the architectural vulnerability of key pipeline components, we propose to harden them in multiple different combinations with varying levels of reliability to cater to the application's requirement while minimizing the power overhead.
The pareto-optimal reliability mode \textbf{\textit{RM7}} is successful in reducing the processor vulnerability by 87\% on average, with area and power overheads of 10\% and 43\%, respectively. 
To further enhance our design space for heterogeneous reliability, we also investigate effective state-compression techniques to reduce the data size of a checkpoint by \textasciitilde 6$\times$.
\changed{Our studies illustrate that in power-constrained scenarios, enabling reliability at a fine granularity, and deploying reliability-heterogeneous super-scalar out-of-order processor bear a significant potential for real-world systems, especially when considering diverse vulnerability profiles of different applications, which can further vary depending upon their input workloads.}


%% file: sections/z_biography.tex
\begin{IEEEbiography}[{\includegraphics[width=1in,height=1.25in,clip,keepaspectratio]{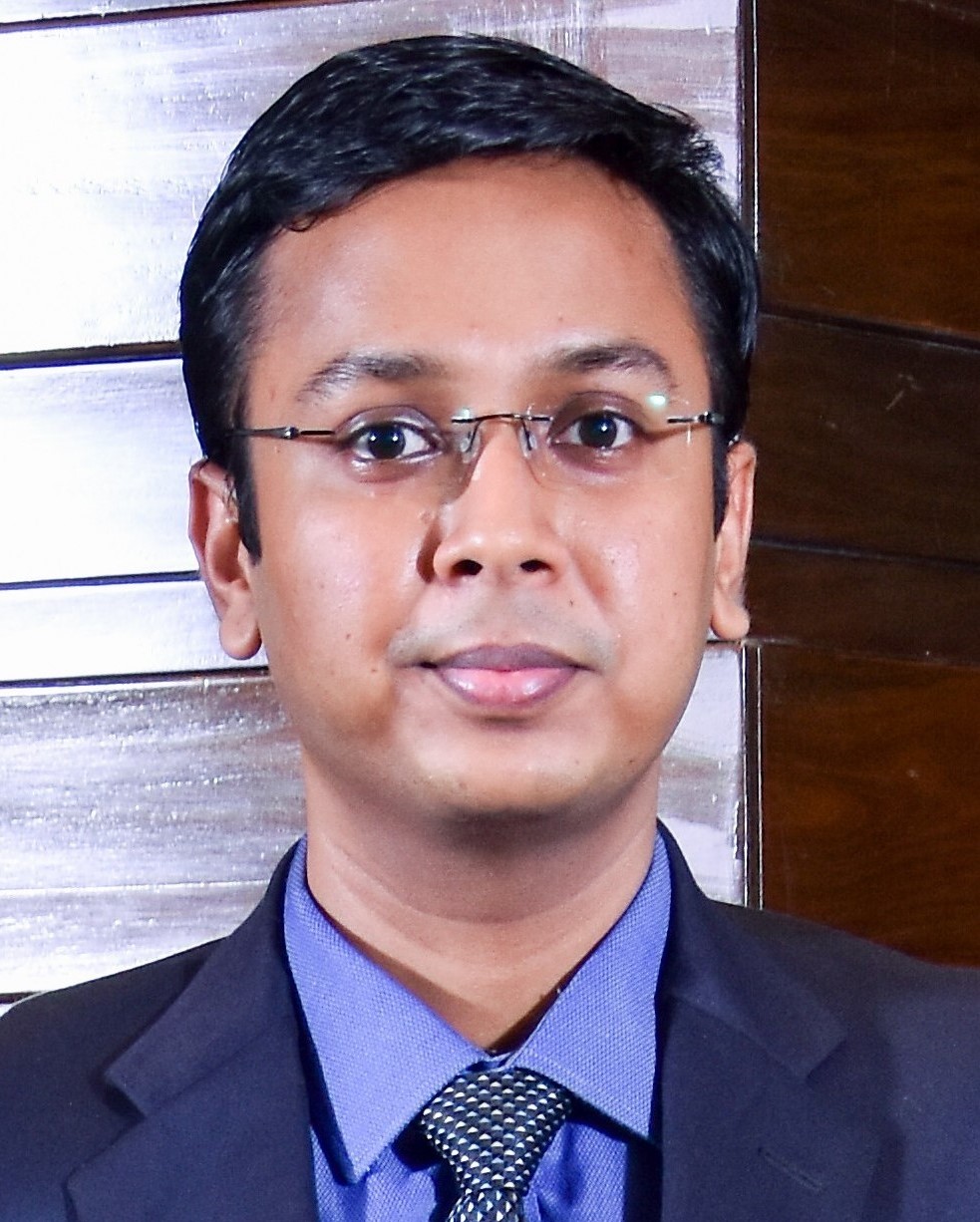}}]{Bharath Srinivas Prabakaran} (S'19) is a PhD Student at the Computer Architecture and Robust Energy-Efficient Technologies (CARE-Tech.) research group, Institute of Computer Engineering, TU Wien, Austria.
He graduated with a Bachelor of Engineering in Electrical and Electronics and a Master of Science in Biological Sciences from the Birla Institute of Technology and Science (BITS), Pilani in 2017. 
He was as a visiting researcher at TU Dresden for a span of 1 year from 2016 to 2017, where he completed his master thesis focused on ``Approximate Computing''.
His research interests include fault-tolerant computing, wearable architectures, healthcare systems, energy-efficient technologies, and embedded machine learning.
\end{IEEEbiography}

\begin{IEEEbiography}[{\includegraphics[width=1in,height=1.25in,clip,keepaspectratio]{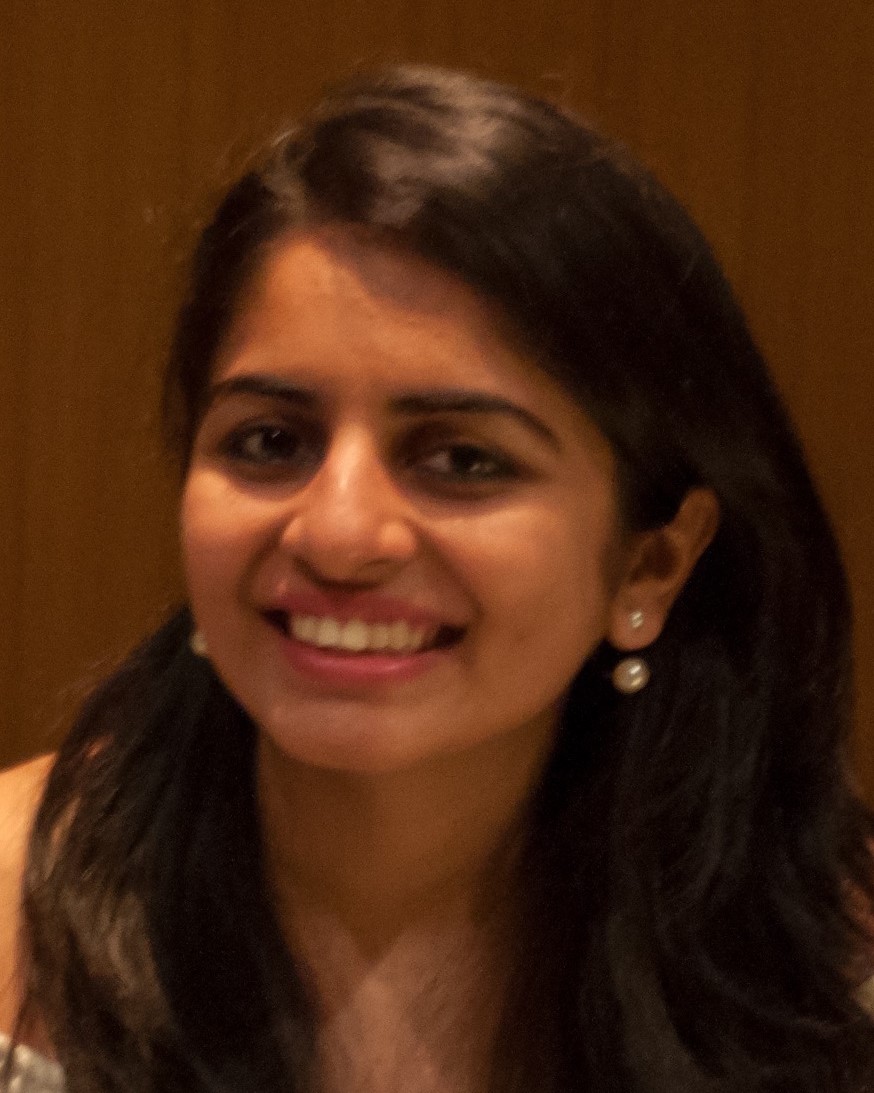}}]{Mihika Dave} is currently working as a software engineer at Facebook, Inc. 
She graduated with a Master of Science in Computer Science from the University of Illinois at Urbana-Champaign with a specialization in natural language processing in 2018. 
In 2016, she graduated with a Bachelor of Engineering from BITS-Pilani, India where she secured the 1$^\text{st}$ rank in the Department of Electrical and Electronics Engineering and received a Bronze Medal in the entire batch of students across all the Science and Engineering Departments.
Her main research interests are heterogeneous fault-tolerance, machine learning, and natural language processing. 
She is the recipient of several scholarships and awards, such as the DAAD-WISE Scholarship, Michal S. Hughes Award in Software Engineering, and BITS-Pilani Merit Scholarship.
\end{IEEEbiography}

\begin{IEEEbiography}[{\includegraphics[width=1in,height=1.25in,clip,keepaspectratio]{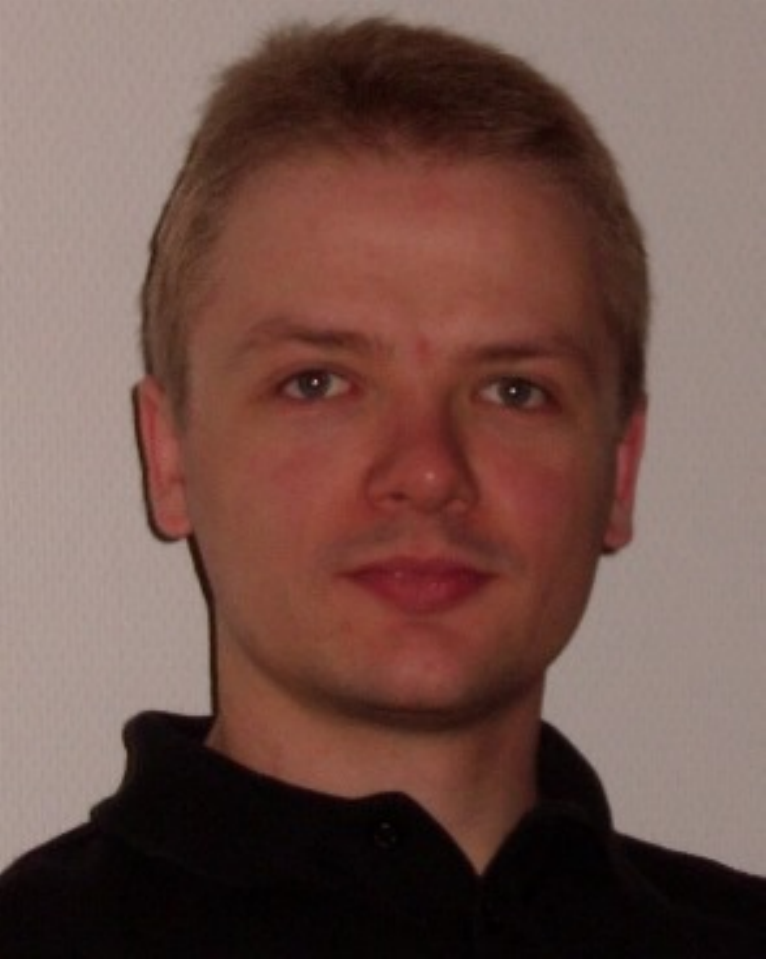}}]{Florian Kriebel} is a university assistant at the Computer Architecture and Robust Energy-Efficient Technologies (CARE-Tech.) research group, Institute of Computer Engineering, TU Wien, Austria. 
He received the M.Sc. degree in computer science from Karlsruhe Institute of Technology (KIT), Germany, in 2013. His current research interests include dependable computing, cross-layer reliability modeling, and optimization. Mr. Kriebel has received the CODES+ISSS 2011 and 2015 Best Paper Awards. 
\end{IEEEbiography}

\begin{IEEEbiography}[{\includegraphics[width=1in,height=1.25in,clip,keepaspectratio]{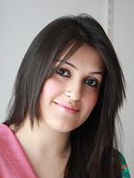}}]{Semeen Rehman} is currently on a Laufbahnstelle (Tenure-Track Assistant Professor) position at Institute of Computer Technology (ICT), Faculty of Electrical Engineering and Information Technology,  Technische Universit{\"a}t Wien (TU Wien). Before that, she was a post-doctoral researcher at the Technische Universit{\"a}t Dresden (TU Dresden) and Karlsruhe Institute of Technology (KIT), Germany since 2015. She received her Ph.D. in computer science on 15.July.2015 from KIT, Germany. Her main research interests are dependable systems, cross-layer design for error resiliency with a focus on run-time adaptations, emerging computing paradigms like approximate computing, hardware security, energy-efficient computing, embedded systems, MPSoCs, IoT and CPS. 
Dr. Rehman has contributed key ideas that have led to various DFG projects such as GetSURE and GetSURE-II at the KIT, which focused on enabling reliability across multiple software and hardware layers.
At the Chair for Processor Design at  Technische Universit{\"a}t Dresden, Germany, she initiated the research on Reconfigurable Approximate Computing. 
Dr. Rehman received the CODES+ISSS 2011 and 2015 Best Paper Awards, DATE 2017 Best Paper Award Nomination, several HiPEAC Paper Awards, Richard Newton Young Student Fellow Award at DAC 2015, and Research Student Award at KIT in 2012. She has served on the TPC of multiple premier conferences on design automation and embedded systems (like DATE and CASES), and has (co-)chaired sessions at the DATE 2017, 2018, and 2019 conferences. She co-authored 1 book, multiple book chapters, and 30+ publications in premier journals and conferences.
\end{IEEEbiography}

\begin{IEEEbiography}[{\includegraphics[width=1in,height=1.25in,clip,keepaspectratio]{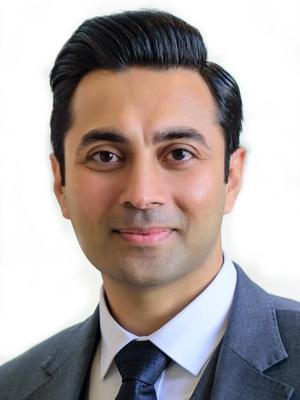}}]{Muhammad Shafique} (M'11 - SM'16) is a full professor of Computer Architecture and Robust Energy-Efficient Technologies (CARE-Tech.) at the Institute of Computer Engineering, TU Wien, Austria since Nov. 2016. 
He received his Ph.D. in Computer Science from Karlsruhe Institute of Technology (KIT), Germany, in Jan.2011. 
Before, he was with Streaming Networks Pvt. Ltd. where he was involved in research and development of video coding systems for several years. His research interests are in computer architecture, power-/energy-efficient systems, robust computing, hardware security, Brain-Inspired computing trends like Neuromorphic and Approximate Computing, hardware and system-level design for Machine Learning and AI, emerging technologies \& nanosystems, FPGAs, MPSoCs, and embedded systems. 
His research has a special focus on cross-layer modeling, design, and optimization of computing and memory systems, as well as their deployment in use cases from Internet-of-Things (IoT), Cyber-Physical Systems (CPS), and ICT for Development (ICT4D) domains.

Dr. Shafique has given several Keynotes, Invited Talks, and Tutorials. 
He has also organized many special sessions at premier venues and served as the Guest Editor for IEEE Design and Test Magazine and IEEE Transactions on Sustainable Computing. 
He has served on the TPC of numerous prestigious IEEE/ACM conferences. 
Dr. Shafique received the 2015 ACM/SIGDA Outstanding New Faculty Award, six gold medals in his educational career, and several best paper awards and nominations at prestigious conferences like CODES+ISSS, DATE, DAC and ICCAD, Best Master Thesis Award, DAC'14 Designer Track Best Poster Award, IEEE Transactions of Computer "Feature Paper of the Month" Awards, and Best Lecturer Award. 
Dr. Shafique holds one US patent and has (co-)authored 6 Books, 10+ Book Chapters, and over 200 papers in premier journals and conferences. He is a senior member of the IEEE and IEEE Signal Processing Society (SPS), and a member of the ACM, SIGARCH, SIGDA, SIGBED, and HIPEAC.
\end{IEEEbiography}